\documentclass[prb,aps,twocolumn,,showpacs,floatfix]{revtex4}

\usepackage{graphicx}
\usepackage{dcolumn}
\usepackage{bm}
\usepackage{subfigure}
\usepackage{psfrag}
\usepackage{amssymb,amsmath}
\newcommand{\figwidth}{4 cm}
\newcommand{\bra}[1]{\left<#1\right|}
\newcommand{\ket}[1]{\left|#1\right>}
\newcommand{\G}{{\mathcal G}}

\begin{document}

\title{Conductance Through Graphene Bends and Polygons}

\author{A. Iyengar$^1$, T. Luo$^1$, H.A. Fertig$^{1,2}$ and L. Brey$^3$}

\affiliation{\centerline {1. Department of Physics, Indiana
University, Bloomington, IN 47405}\\
{\centerline{2. Department of Physics, Technion, Haifa 32000, Israel}}
\\ 3. Instituto de Ciencia de Materiales de
Madrid (CSIC),~Cantoblanco,~28049~Madrid,~Spain}
\

\begin{abstract}
We investigate the transmission of electrons between conducting
nanoribbon leads oriented at multiples of 60$^{\circ}$ with
respect to one another, connected either directly or through
graphene polygons. A mode-matching analysis suggests that the
transmission at low-energies is sensitive to the precise way in
which the ribbons are joined. Most strikingly, we find that
armchair leads forming 120$^{\circ}$ angles can support either a
large transmission or a highly suppressed transmission, depending
on the specific geometry.  Tight-binding calculations demonstrate
the effects in detail, and are also used to study transmission at
higher energies as well as for zigzag ribbon leads.
\end{abstract}


\maketitle

\section{Introduction}
Graphene, a two-dimensional honeycomb
lattice of carbon, is one of the most interesting new
low-dimensional materials to have become available
in the laboratory in the last few years
\cite{novoselov}.  When undoped, the low energy
physics of this system is dominated by two Dirac
points \cite{review}.  The wavefunctions associated
with states near them are described by  spinors,
whose amplitudes represent the probability density
to find electrons on either of the two honeycomb sublattices.
Because the fermion spectrum is gapless, these
spinors
have well-defined helicity, leading to an absence of
backscattering from impurities \cite{review2,shon}.
The observed metallic behavior of undoped graphene
is likely to be a manifestation of this suppressed
backscattering \cite{novoselov}.

Among the interesting and potentially useful properties
of graphene is the prospect of ``tailoring'' its electronic
properties by cutting it into ribbons of well-defined
widths along various symmetry directions \cite{review,brey1,brey2,brey3}.
Recent experimental work \cite{han,chen,li}
has confirmed the possibility of tuning transport
gaps of graphene ribbons via their widths, although the
quality of these ribbons is not yet high enough to be usefully compared
with the expected \cite{brey1} width
dependence of ideal ribbons.  In applications
one generically needs to join such ribbons
together as interconnecting wires or elements of
a device.  Understanding the transport through such
junctions is then important in designing graphene
geometries with desirable behaviors.  This the subject
of our study.

Changing the direction of electron currents
would presumably be an important aspect of any
graphene-based circuit.
In both quantum wires and electromagnetic waveguides it is
known \cite{londergan} that the transmission
through bends depends on the detailed nature of the bend
geometry.
This raises the prospect
that the conductive behavior of a graphene junction may differ
substantially from the properties of its individual nanoribbon leads.

In the simplest situation, the nanoribbon leads meeting at a junction
are of the same type (armchair or zigzag), restricting
the bend angles to either 60$^{\circ}$ or 120$^{\circ}$.  The latter case
is particularly interesting due to the behavior of the low energy
states near the Dirac points under 60$^{\circ}$
rotation.
These states may be constructed, within the
${\bf k} \cdot {\bf p}$ approximation, from products
of the exact wavefunctions at the Dirac points, which
vary rapidly in real space,
with slowly varying envelope functions \cite{review2}.
The rotation induces a transformation on the fast component of the
wavefunction which exchanges both the valleys and the sublattices.
As a result, states near the Dirac points are nearly orthogonal to
their 60 degree rotation in a confined geometry.  Since a ribbon
with a 120$^{\circ}$ junction may be viewed as 60$^{\circ}$
deflection of the electron trajectory, one might expect a
generically suppressed transmission through 120$^{\circ}$ bends.

Our studies show that, while this na\"ive reasoning
is sometimes borne out,
in general the transmission through such bends is  not
universal and depends critically on the details of the junction.
In this paper,
we focus on geometries in which 120$^{\circ}$ bends in armchair
nanoribbons are realized either by a ``kink'', or by attachment
to triangular or hexagonal central regions.
We focus mostly on the energy region  $|E_F| < E_e$, $E_e$ being
the band edge of the first excited band of the nanoribbon leads,
where there is only one channel available for conduction.
In this low energy region, we know that the eigenstates
of graphene nanoribbons may be understood within a continuum
approximation (the Dirac equation) when appropriate boundary
conditions are adopted \cite{brey1}, so that a description
involving matching of these wavefunctions at junctions becomes possible.
As we will discuss in detail, these geometries produce
quite different transmission behavior at low energies. This
suggests the prospect of using different types of junctions to
tailor transmission through a set of graphene nanoribbons.

To understand why the geometry is crucial, at low energies we may adopt
wavefunctions for
armchair ribbons obtained from solutions
to the Dirac equation \cite{brey1}, and consider how
they might be appropriately matched at a junction.
We focus on armchair nanoribbons whose
widths are chosen so that they are metallic \cite{brey1}.
A ``mode-matching'' procedure
may be formally developed \cite{londergan} to compute
conductance properties of the system.  This becomes
particularly simple when only the lowest energy
transverse modes are retained -- the ``single-mode approximation'' (SMA).

\begin{figure}[htb]
\centering
\includegraphics[width=7.cm]{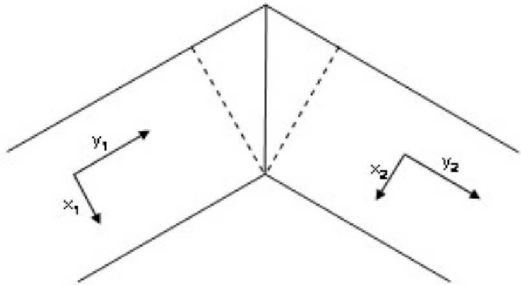}
\caption{
Schematic diagram of a 120$^{\circ}$ junction, illustrating the
coordinate systems for the two ribbons [$(x_1,y_1)$ and $(x_2,y_2)$].
The surface at which the two nanoribbons are joined is the solid
vertical line.
}
\label{fig:simple_AC_junction}
\end{figure}

\begin{figure}[htb]
\includegraphics[width=8.cm]{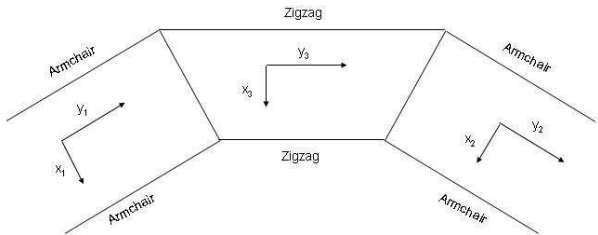}
\caption{Schematic diagram of a 120$^{\circ}$ armchair junction, in which
electrons pass through a length of zigzag ribbon.
Coordinates for the three sections illustrated.
Wavefunctions must be matched along the two solid lines
connecting the armchair ribbons to the zigzag ribbon.
}
\label{fig:simple_ZZ_junction}
\end{figure}

In the simplest case, the 120$^{\circ}$ junction
(see Fig. \ref{fig:simple_AC_junction}),
we shall see that the in the zero energy limit, incoming and
outgoing modes can be matched up perfectly at the junction, so
that at low energies the electrons are {\it maximally}
transmitted. By contrast, passage through a short length of zigzag
ribbon (Fig. \ref{fig:simple_ZZ_junction})
involves intermediate transverse states that are
strongly localized to the edges, which become
orthogonal to those in the armchair leads very close
to zero energy.  This implies blocked transmission in a narrow range
of energies near zero.  Transmission through an equilateral triangle
with two attached armchair leads may be viewed as a special case of this class of geometries.

\begin{figure}[htb]
\centering
\includegraphics[width=7.cm]{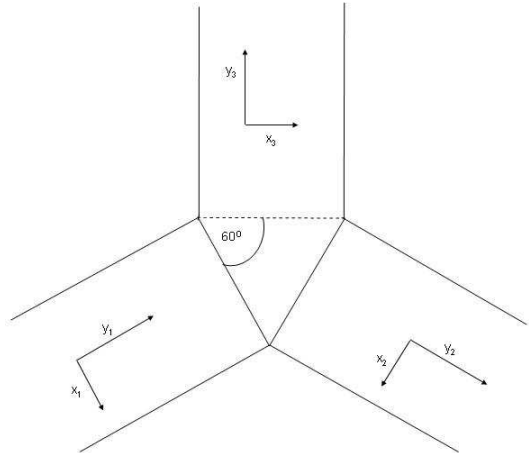}
\caption{
Schematic diagram of an equilateral triangle
with three armchair leads, in which the triangle
is taken as an ``endcap'' for the vertical lead.
Coordinates for the three sections illustrated.
Wavefunctions must be matched along the two solid lines
of the triangle.
}
\label{fig:simple_EQT_junction}
\end{figure}

A closely related geometry is transmission through an
equilateral triangle with {\it three} leads.
Here the system may be constructed from three appropriately cut armchair leads
as illustrated in Fig. \ref{fig:simple_EQT_junction}.  As we will show
in detail for this case, rapid oscillation of the fast component of the 
wavefunction makes the transmission 
very sensitive to the precise way in which these
armchair leads are
connected to the triangle, so that
one may obtain either large or vanishingly small
conductance in the low energy limit.  This sensitivity to the
geometry is ubiquitous for such graphene
nanostructures, suggesting that a wide variety of conductance
properties can in principle be engineered into very similar
geometries.

The SMA thus leads one to expect qualitatively different conductances
for 120$^{\circ}$ armchair junctions at low energies, depending
on precisely how they are joined.  We have tested these
expectations using a tight-binding model \cite{datta}
for the ribbons and the regions joining them.
Again, the simplest case is
the
120$^{\circ}$ junction between two armchair nanoribbons.
An example of a specific connecting geometry and
its associated low-energy conductance,
computed in a tight-binding model 
for ribbons of conducting width, is illustrated
in Fig. \ref{fig:120Bend_AC}.  In agreement with the SMA result,
one obtains almost perfect transmission at low energies,
nearly to the bottom of the first excited transverse subband energy.

\begin{psfrags}
\psfrag{xlabel}[][][0.7]{$E_F/E_e$}
\psfrag{ylabel}[b][cl][0.7]{G $(e^2/h)$}
\begin{figure}[htb]
\centering
\subfigure[$N$=5]{
\includegraphics[width=\figwidth]{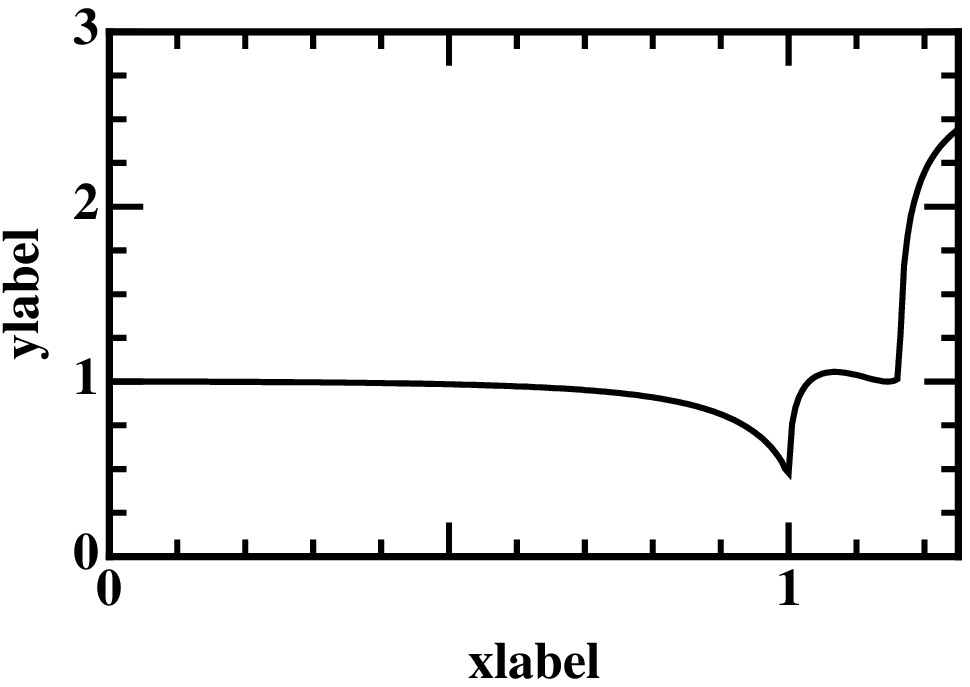}
}
\subfigure[$N$=8]{
\includegraphics[width=\figwidth]{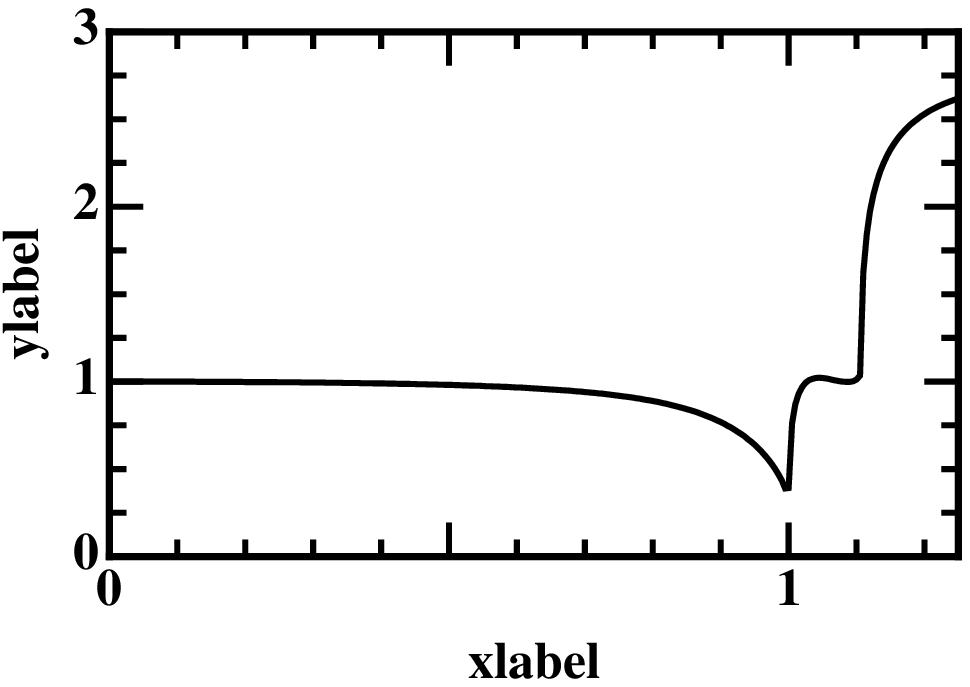}
}
\\
\subfigure[$N$=11]{
\includegraphics[width=\figwidth]{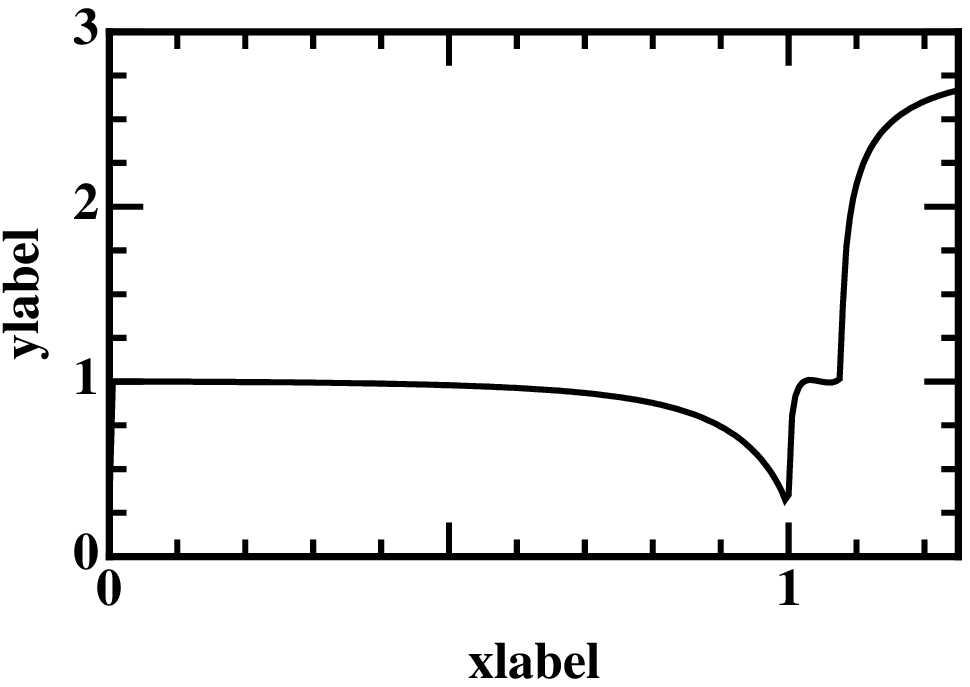}
}
\subfigure[Geometry for $N=8$]{
\includegraphics[width=\figwidth]{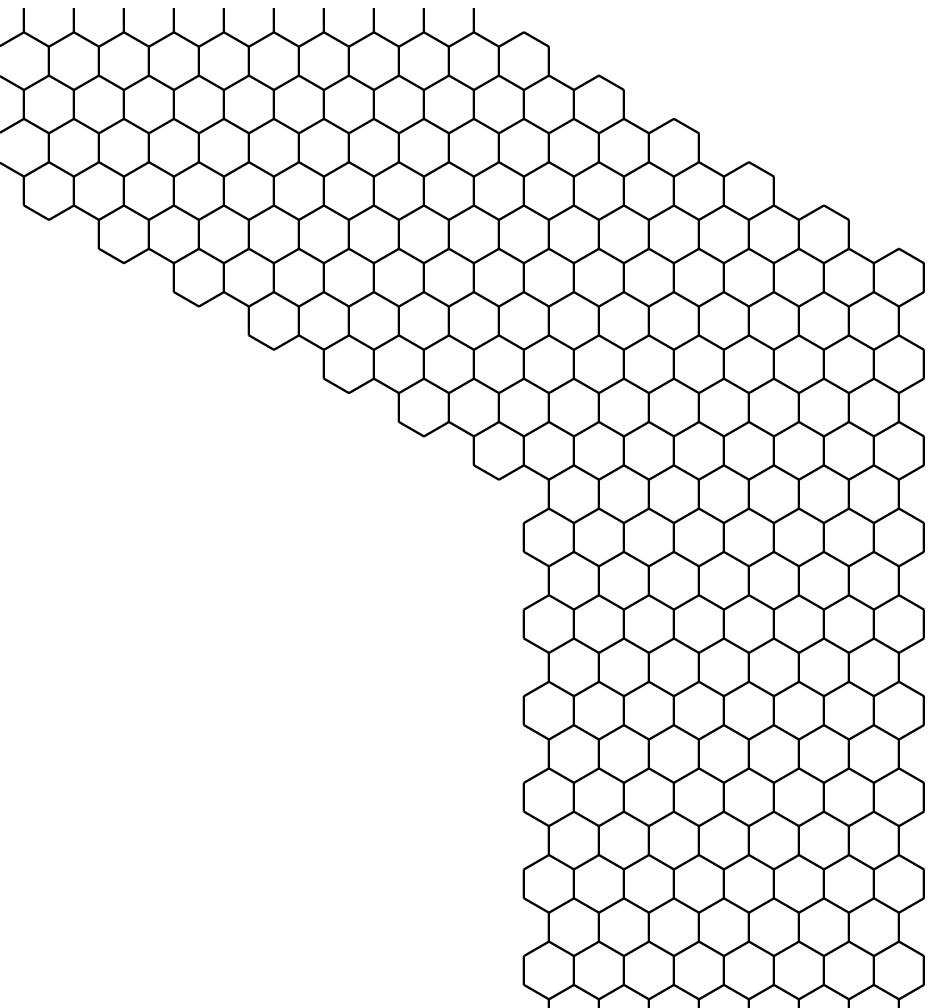}
}
\caption{Conductance per spin of 120-degree bends in armchair nanoribbons having N transverse channels.
Geometry for $N$=8 illustrated in (d).}
\label{fig:120Bend_AC}
\end{figure}
\end{psfrags}

Fig. \ref{fig:Triangle_2Lead_AC} illustrates a typical result for
two leads joined through an equilateral triangle.  As suggested
by the SMA, the conductance is now suppressed near zero energy,
but only over a very narrow range.  Tight-binding studies of
analogous geometries in which armchair leads are
connected by short zigzag ribbons give similar results
at low energies.  One may also compute the conductance
through 120$^{\circ}$ when a third lead is added to the
equilateral triangle.  The results of such calculations
are illustrated in Figs. \ref{fig:Triangle_3Lead_AC_2}
and \ref{fig:Triangle_3Lead_AC_1}.  In the former case,
for which the leads are attached to the triangle at their
widest cross-sectional widths,
one finds a highly suppressed conductance at low energy.
The latter case has the leads attached to the triangle at their
narrow cross-section, and the resulting conductance is large
at low energy.  We shall see both these results may be
understood
from the SMA, as special cases of the geometry
illustrated in Fig. \ref{fig:simple_EQT_junction}.
Thus, the mode-matching
procedure appears to offer a useful framework for
understanding conductance in such geometries.

\begin{psfrags}
\psfrag{xlabel}[][][0.7]{$E_F/E_e$}
\psfrag{ylabel}[b][cl][0.7]{G $(e^2/h)$}
\begin{figure}[htb]
\centering
\subfigure[$N$=8]{
\includegraphics[width=\figwidth]{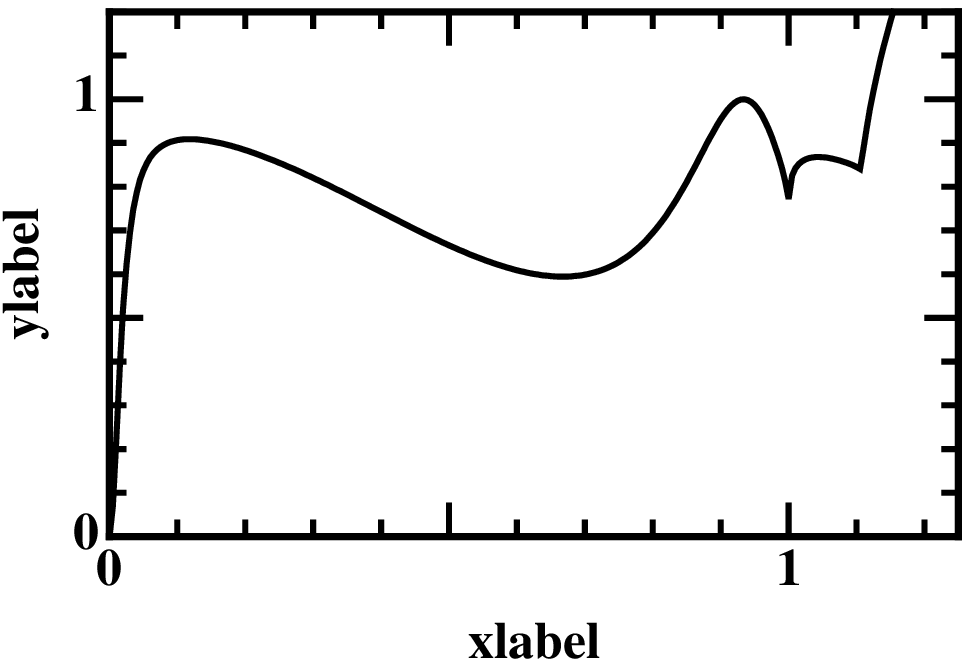}
}
\subfigure[$N$=11]{
\includegraphics[width=\figwidth]{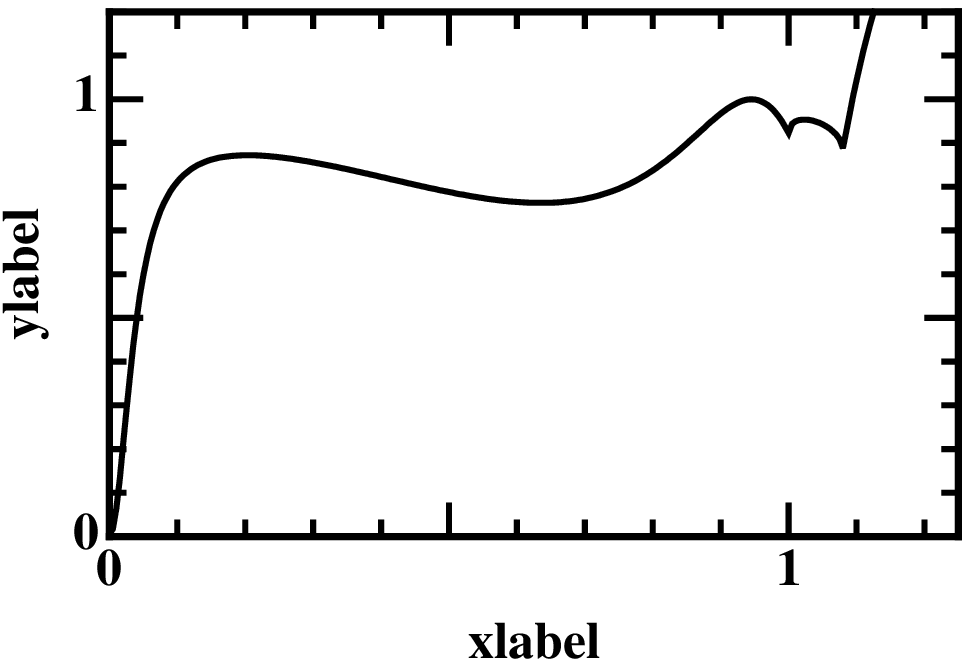}
}
\\
\subfigure[$N$=14]{
\includegraphics[width=\figwidth]{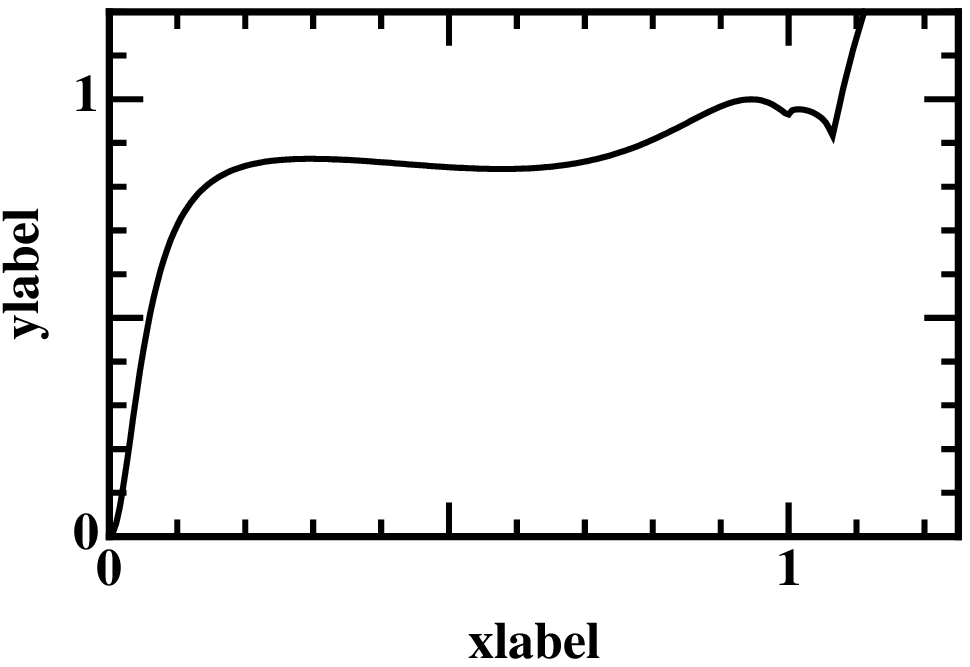}
}
\subfigure[Geometry for $N=8$]
{
\includegraphics[width=\figwidth]{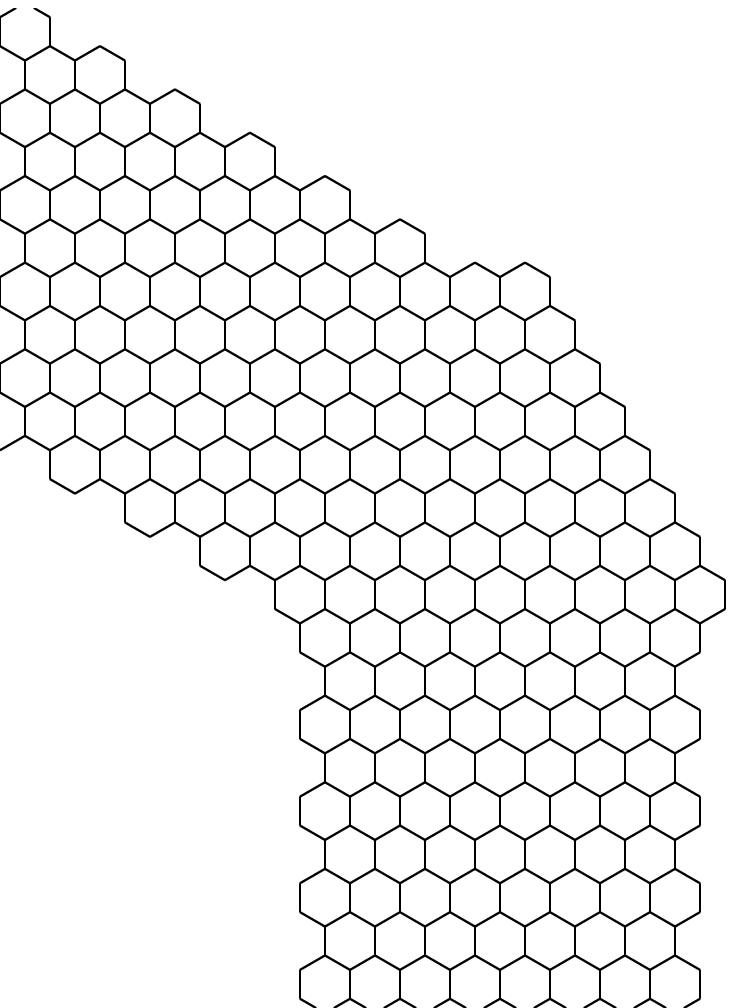}
\label{fig:Triangle_2Lead_AC_Diagram}
}
\caption{Conductance per spin of triangles with 2 armchair nanoribbons leads having N
transverse channels. Geometry for $N$=8 illustrated in (d).
}
\label{fig:Triangle_2Lead_AC}
\end{figure}
\end{psfrags}


\begin{psfrags}
\psfrag{xlabel}[][][0.7]{$E_F/E_e$}
\psfrag{ylabel}[b][cl][0.7]{G $(e^2/h)$}
\begin{figure}[t]
\centering
\subfigure[$N$=5]{
\includegraphics[width=\figwidth]{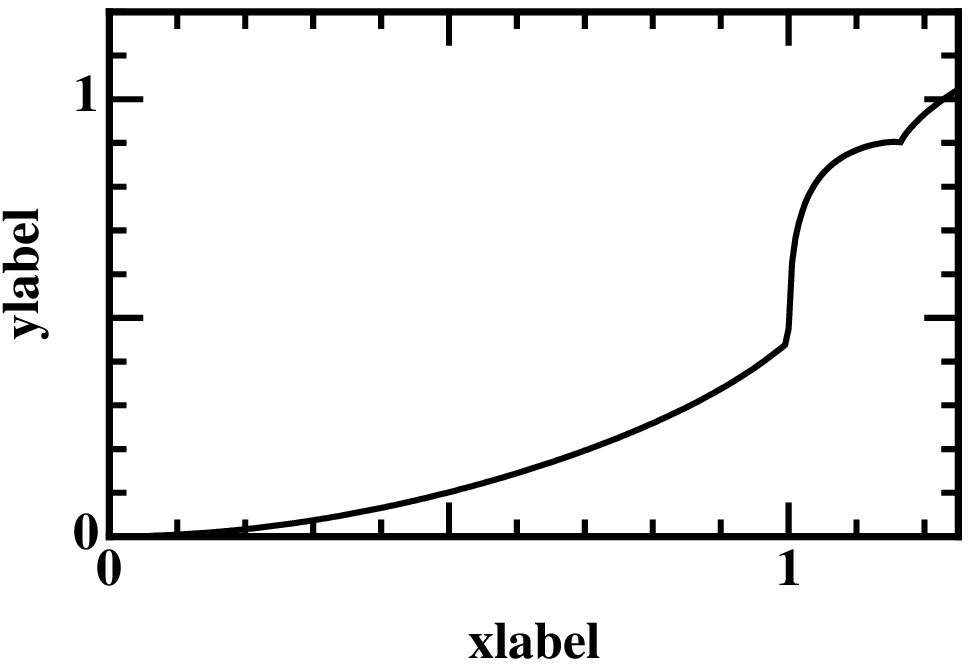}
}
\subfigure[$N$=8]{
\includegraphics[width=\figwidth]{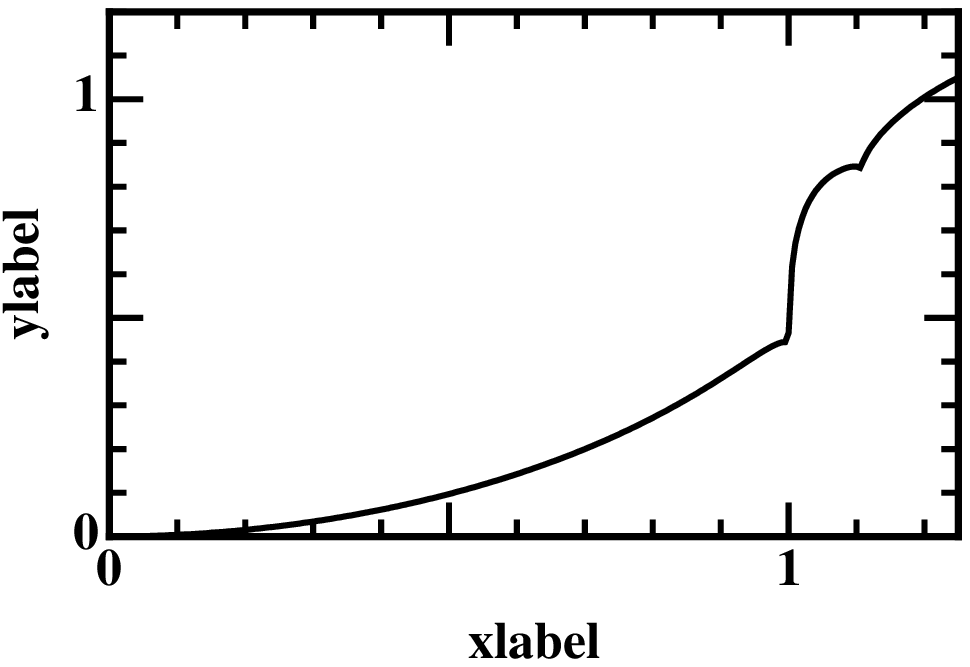}
}
\\
\subfigure[$N$=11]{
\includegraphics[width=\figwidth]{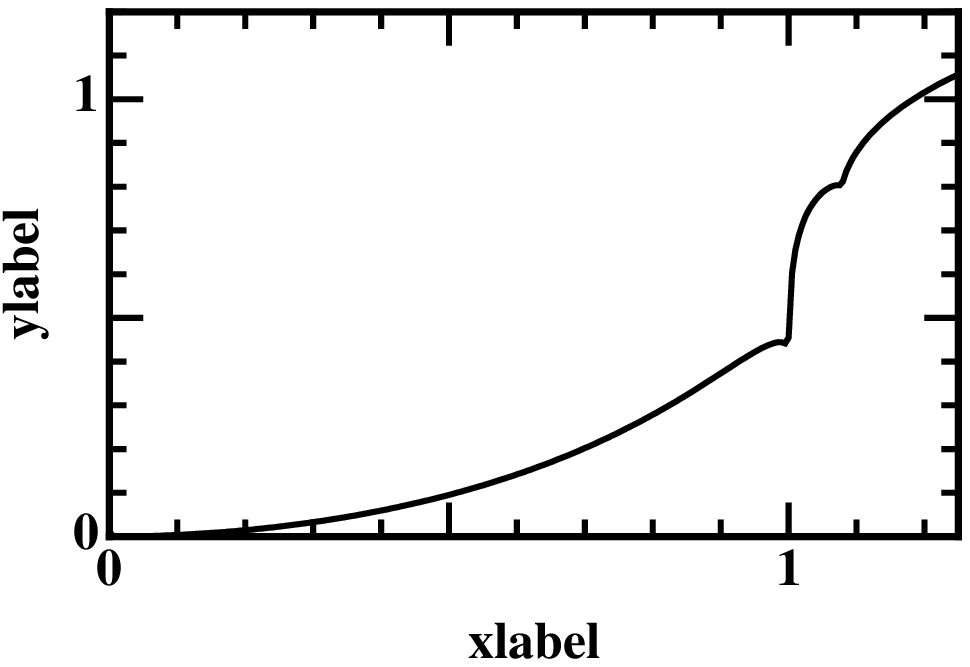}
}
\subfigure[Geometry for $N=8$]
{
\includegraphics[width=\figwidth]{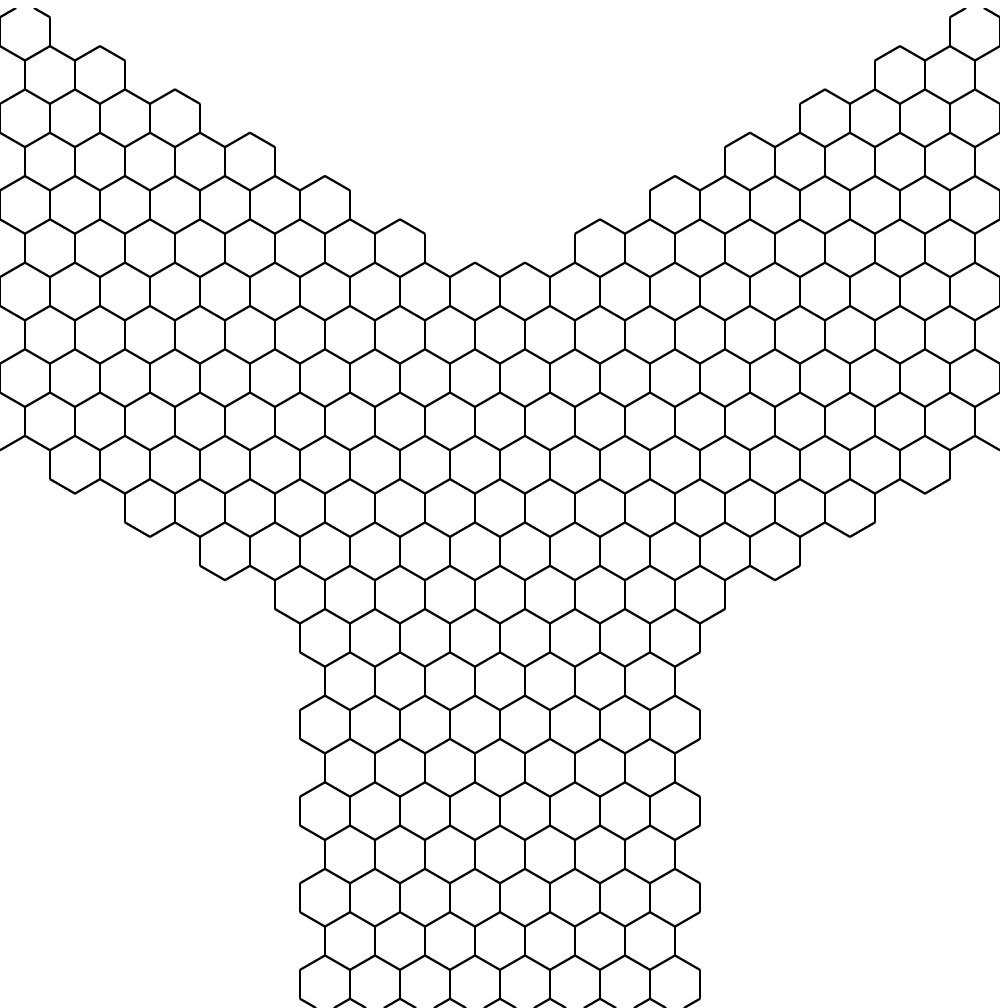}
}
\caption{Conductance per spin of triangles with 3 armchair nanoribbon leads having N transverse channels,
with ribbons connecting to the triangle along their wider cross-section.
Geometry for $N$=8 illustrated in (d).
}
\label{fig:Triangle_3Lead_AC_2}
\end{figure}
\end{psfrags}

\begin{psfrags}
\psfrag{xlabel}[][][0.7]{$E_F/E_e$}
\psfrag{ylabel}[b][cl][0.7]{G $(e^2/h)$}
\begin{figure}[t]
\centering
\subfigure[$N$=5]{
\includegraphics[width=\figwidth]{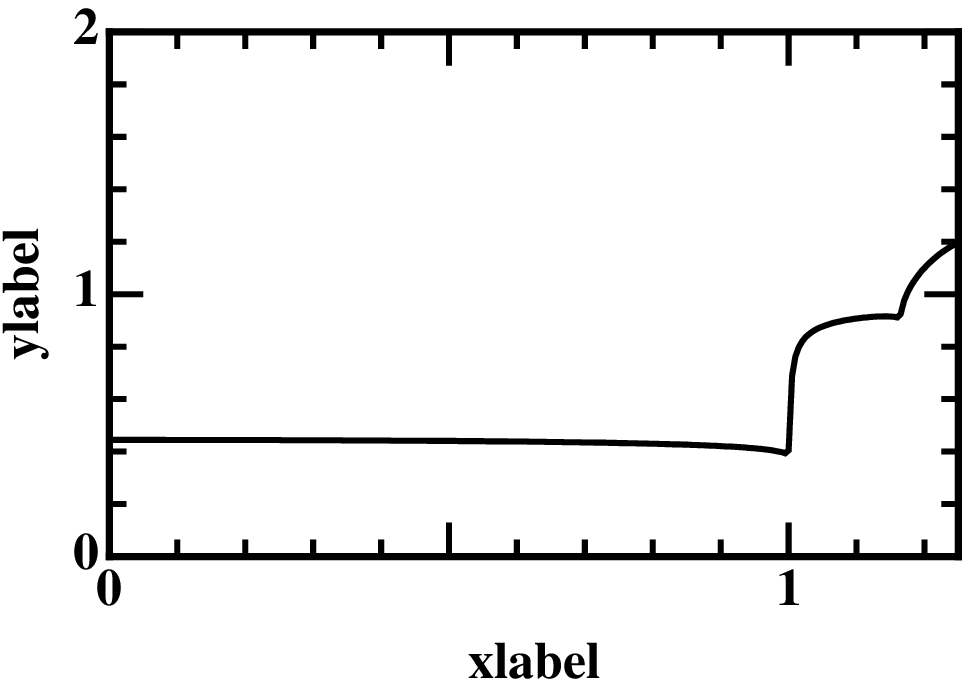}
}
\subfigure[$N$=8]{
\includegraphics[width=\figwidth]{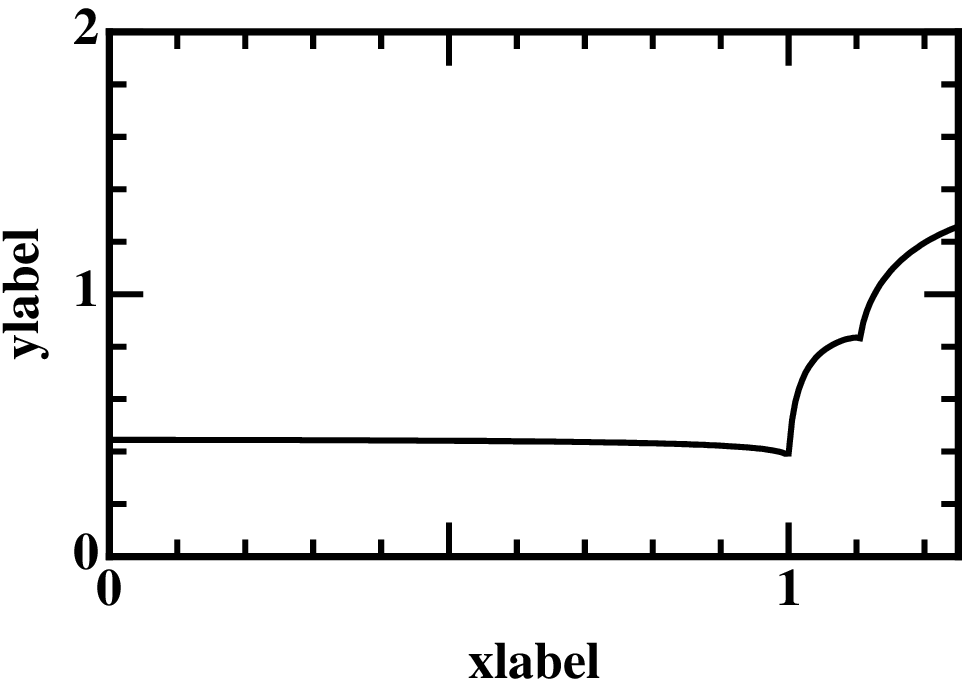}
}
\\
\subfigure[$N$=11]{
\includegraphics[width=\figwidth]{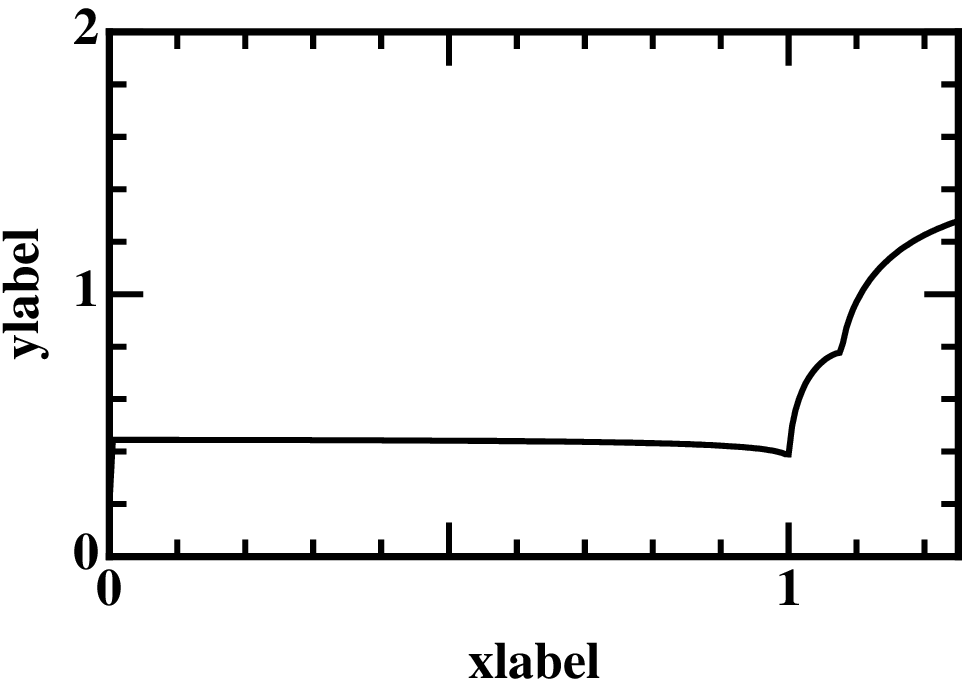}
}
\subfigure[Geometry for $N=8$]
{
\includegraphics[width=\figwidth]{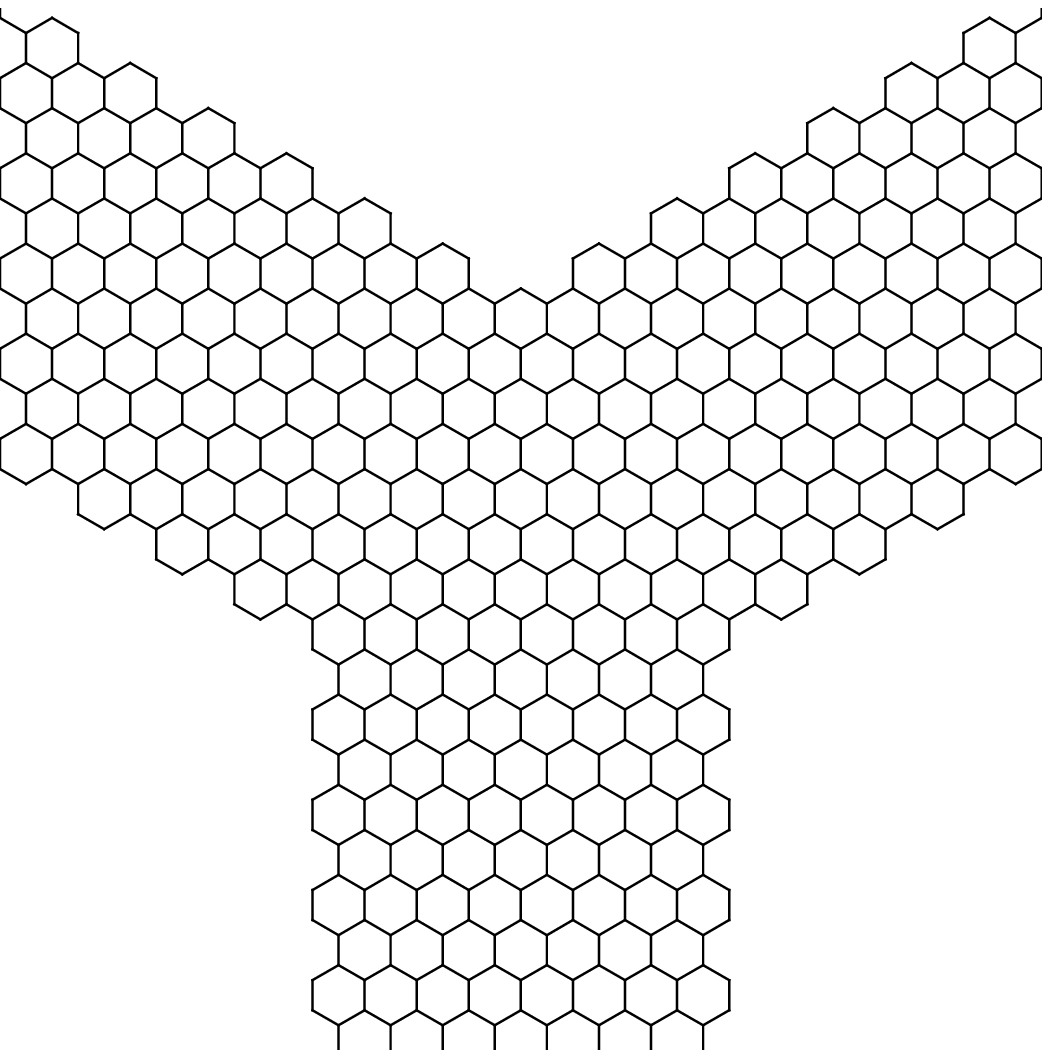}
}
\caption{Conductance per spin of triangles with 3 armchair nanoribbon leads having N transverse channels,
with ribbons connecting to the triangle along their narrower cross-section.
Geometry for $N$=8 illustrated in (d).
}
\label{fig:Triangle_3Lead_AC_1}
\end{figure}
\end{psfrags}

The tight-binding calculations also allow us to
examine the conductance at higher energies, as well
as to consider other geometries.  Here we summarize
a few results from such studies, and give further
details later on.
(1) Transmission studies through hexagons with armchair leads reveal
low-energy conductances that are also suppressed at low
energies for 120$^{\circ}$ transmission.  However, in this
case the suppression is quite dramatic throughout the lowest
subband.
Conductances per spin through 60$^{\circ}$ and 180$^{\circ}$ is
a finite fraction of $e^2/h$ in the lowest subband.
(2) At higher
energies, we find in all of the geometries
considered that the transmission as a function
of energy approximately follows the density of
states for bulk graphene, provided the ribbon widths
are large enough.
However, very close to the energy of the van Hove singularity we
find a strong suppression or enhancement of the conductance
depending on the angle between the leads.
(3) 120$^{\circ}$ zigzag ribbon junctions have a more complicated
evolution with increasing ribbon width than their
armchair cousins.  Most notably, very close to zero energy the conductance
oscillates between large and small values as the ribbon
width is incremented by a single unit, reminiscent of recent
results for $p-n$ junctions of zigzag ribbons \cite{akhmerov}.

The remainder of this article is organized as follows.
In Section II, we discuss the mode matching analysis
in more detail, and explain how the SMA leads to
the expectations described above.  Section III
is devoted to describing
the numerical methods used to compute the
conductance through the various geometries in the tight-binding
model.  In Section IV we provide further details of
our numerical results.  Finally,
we conclude in
Section V with a summary of our results.

\section{Mode-Matching Analysis}

The conduction properties of electron systems in which current
is injected into and removed from a region with a defined
shape and potential,
through infinite leads with known cross-section,  can
be understood by exploiting their analogy with electromagnetic
waveguides \cite{londergan}.  A conceptually simple
approach to their analysis is to divide the system
into components where the wavefunctions may be
computed with appropriate boundary conditions, and
then ``stitch'' the wavefunctions together at the
boundaries by matching their amplitudes and, in the
case of wavefunctions controlled by the Schroedinger equation,
their derivatives.
Such calculations are made analytically tractable
by employing the single-mode approximation (SMA),
in which only the lowest subband of the external
leads is retained, which produces qualitatively and
often quantitatively good results provided one
works away from scattering resonances of the system \cite{londergan}.
A similar strategy may be employed to understand our low energy numerical
results for systems with armchair leads.

The simplest case to consider is the 120$^{\circ}$ junction between
two armchair ribbons illustrated in Fig. \ref{fig:simple_AC_junction}. 
The geometry is divided into regions 1 and 2 with corresponding 
wavefunctions $\psi^{(1)}$ and $\psi^{(2)}$ and coordinate systems
$(x_1,y_1)$, $(x_2,y_2)$. 
At low energies one may write down
approximate forms for the ribbon eigenstates \cite{brey1}.
For momentum ${\bf p}$
these have the form
\begin{widetext}
\begin{equation}
\psi^{(i)}_{p_x,p_y}(x_i,y_i)={1 \over {\sqrt{2W}}}\left\{ \left(
\begin{array}{c} 1 \\
\frac{p_x+ip_y}{p}
\end{array}
\right)e^{i{\bf K}\cdot{\bf r_i}}e^{ip_xx_i}
-
\left(
\begin{array}{c} 1 \\
\frac{p_x+ip_y}{p}
\end{array}
\right)e^{i{\bf K}^{\prime}\cdot{\bf r}_i}e^{-ip_xx_i}
\right\}
e^{ip_yy_i},
\label{arm_ribbon}
\end{equation}
\end{widetext}
where the upper (lower) entry represents the amplitude on the $A$ ($B$)
sublattice, $W$ is the ribbon width, the ${\bf r}_i=(x_i,y_i)$
represent the positions of lattice points 
on the two ribbons, ${\bf K}=(-{{4\pi} \over {3a}},0)$,
${\bf K}^{\prime}=({{4\pi} \over {3a}},0)$, and $a$ is the lattice
constant.  The value of $p_x$ must be chosen such that the total
amplitude at the edges of the ribbons vanishes, hence $p_x \rightarrow p_n$
comes in quantized values \cite{brey1}.  For metallic ribbons, 
the lowest subband
satisfies $p_{n=0}=0$.  These wavefunctions have energy
$\varepsilon=v_F|{\bf p}|$, where $v_F$ is the speed of electrons
near the Dirac points.  Due to particle-hole symmetry, we
restrict our analysis to $\varepsilon \geq 0$, setting $\varepsilon=E_F$
for the determination of the conductance at zero temperature. 

A general wavefunction in which current is injected only from
the left in the lowest subband ($0<\varepsilon < v_F p_1$) may be written in the form
\begin{eqnarray}
\psi^{(1)} &=& A_0 \psi_{p_n=0,p_y}^{(1)} + \sum_n B^{(1)}_n \psi_{p_n,-p_y}^{(1)}
\label{mode_expand1}\\
\psi^{(2)} &=& \sum_n B^{(2)}_n \psi_{p_n,p_y}^{(2)}.
\label{mode_expand2}
\end{eqnarray}
In this expansion, it is implicitly understood that for modes where $v_Fp_n>\varepsilon$,
one replaces $ip_y$ with $\pm \kappa_y$ in Eq. \ref{arm_ribbon} such that
the wavefunctions appearing in Eqs. \ref{mode_expand1} and \ref{mode_expand2}
are evanescent
as one moves away from the junction.  The $B^{(i)}$ coefficients are determined
in terms of $A_0$ by matching the wavefunctions, for both sublattices,
on the solid line shown in Fig. \ref{fig:simple_AC_junction}.  The conductance per spin is then given by
$G={{e^2} \over h} (|B_0^{(2)}|^2/|A_0|^2).$  

\begin{figure}
\centering
\includegraphics[width=\columnwidth]{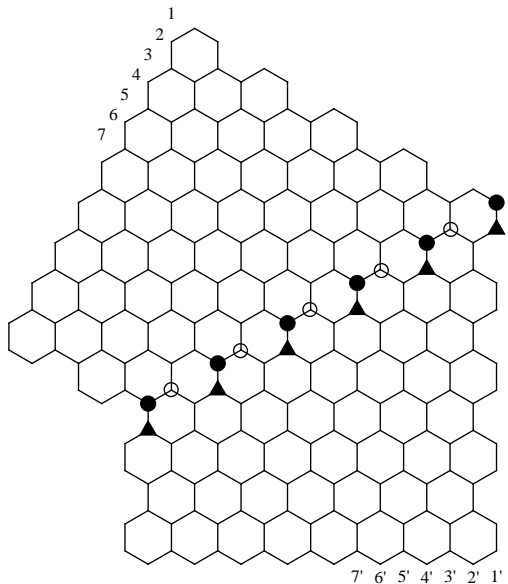}
\caption{Diagram detailing the joining surface between leads
for the simple 120$^{\circ}$ bend, as well as the labelling 
scheme for $x$ and $x'$ coordinates. Matching the wavefunction
on the dark circles and triangles accounts for current continuity
across the joining surface.
}
\label{fig:120degree_bend_join}
\end{figure}

To carry out this procedure, one
must first specify a set of matching conditions at the joining
surface that guarantees continuity of the wavefunctions and
the current across the junction.  One possible choice is illustrated
in Fig. \ref{fig:120degree_bend_join}, where it is now convenient
to change notation slightly and refer to wavefunctions $\psi,\psi'$ and  
coordinates $(x,y), (x',y')$ in regions 1 and 2, respectively.
Equating $\psi$ to $\psi'$ on the
joining line (open and closed circles) guarantees continuity of
the wavefunctions.  Matching currents across these junction points
can be more complicated because this in general involves products of
wavefunctions on either side of a bond.  However we can greatly simplify
the latter matching condition by anticipating the SMA, for which
we will use the wavefunctions
of Eq. \ref{arm_ribbon}, which vanish on the open circles in Fig. \ref{fig:120degree_bend_join}
for the lowest subband.
Thus one only need match the currents on the bonds connecting the closed
circles to the triangles.  This may be accomplished straightforwardly
by matching the wavefunctions on the triangles as well.  We focus on
the zero energy transmission and take the $p_y \rightarrow 0$ limit.
The resulting matching conditions may now be
written explicitly in the form
\begin{equation}
\begin{split}
&\psi^{(A)}_{0,0}(x=1,y)=\psi^{(B)\prime}_{0,0}(x^{\prime}=1^{\prime},y^{\prime}), \\
&\quad \psi^{(B)}_{0,0}(x=2,y)=\psi^{(A)\prime}_{0,0}(x^{\prime}=1^{\prime},y^{\prime}), \\
&\qquad \psi^{(B)}_{0,0}(x=3,y)=\psi^{(A)\prime}_{0,0}(x^{\prime}=3^{\prime},y^{\prime})\\
&\psi^{(A)}_{0,0}(x=4,y)=\psi^{(B)\prime}_{0,0}(x^{\prime}=4^{\prime},y^{\prime}), \\
&\quad \psi^{(B)}_{0,0}(x=5,y)=\psi^{(A)\prime}_{0,0}(x^{\prime}=4^{\prime},y^{\prime}), \\
&\qquad \psi^{(B)}_{0,0}(x=6,y)=\psi^{(A)\prime}_{0,0}(x^{\prime}=6^{\prime},y^{\prime})\\
&\psi^{(A)}_{0,0}(x=7,y)=\psi^{(B)\prime}_{0,0}(x^{\prime}=7^{\prime},y^{\prime}), \\
&\quad \psi^{(B)}_{0,0}(x=8,y)=\psi^{(A)\prime}_{0,0}(x^{\prime}=7^{\prime},y^{\prime}),\\
&\qquad \psi^{(B)}_{0,0}(x=9,y)=\psi^{(A)\prime}_{0,0}(x^{\prime}=9^{\prime},y^{\prime})\\
&\qquad \qquad \qquad \qquad \vdots
\end{split}
\label{match-armchair}
\end{equation}
where the $x$ and $x'$ labels are defined in 
Fig. \ref{fig:120degree_bend_join}.
Simple phase factors may be added to these matching conditions to generalize them
for the case $p_y \ne 0$.
Note that in writing these equations we have equated $A$ sites
of the left lead with $B$ sites
of the right.  This may be understood if one constructs the
junction by starting with a single ribbon, excises
an equilateral triangular region from the center, 
and joins the two resulting ribbons to form the
junction.  In doing this one finds that at the
junction, $A$ sites are indeed matched up to $B$ sites.
Note that no explicit interchange of the {\bf K} and ${\bf K^{\prime}}$
valleys is needed (as is the case in $60^{\circ}$ rotations)
because the real space coordinate system has also
been rotated.

Eqs. \ref{match-armchair} are a realization of setting Eq. \ref{mode_expand1}
to Eq. \ref{mode_expand2} on the joining surface.
To proceed, we wish to represent the wavefunctions on the matching
points in an expansion in terms of wavefunctions of the form in
Eq. \ref{arm_ribbon}.
Formally, this is accomplished by
multiplying these equations
by $\psi^{(2)*}_{p_n^{\prime},p_y^{\prime}}(x,y)$, and then
integrating $(x,y)$ on the joining surface, i.e., summing
over the points where the wavefunctions have been matched.
(Note $p_y^{\prime}$ is chosen
such that $v_F^2(p_n^{\prime 2}+p_y^{\prime 2})=\varepsilon^2$.)
This results in
a set of equations relating the $A_0$ and $B^{(i)}$ amplitudes.
A second set of equations may be generated by multiplying the matching equation
by $\psi^{(1)*}_{p_n^{\prime},p_y^{\prime}}(x,y)$ and integrating
on the joining surface.
The two sets form an in principle
infinite dimensional matrix equation that relates the $B^{(i)}$ coefficients to $A_0$.

Carrying out this procedure is vastly simplified by adopting the
SMA, in which only the lowest transverse
mode, $p_n=0$, is retained in the matrix equation \cite{londergan}.
To demonstrate the perfect transmission at low energy in the
junction illustrated in Fig. \ref{fig:simple_AC_junction},
it is convenient to consider the {\it reflection}
amplitude in the SMA.
This is proportional to the integral
\begin{equation}
M_{0,0}(p_y) \equiv \int d\lambda \,\, \psi^{(1)*}_{0,-p_y}(x(\lambda),y(\lambda)) \,
\psi^{(1)}_{0,p_y}(x(\lambda),y(\lambda)),
\label{M00}
\end{equation}
where $\lambda$ parameterizes the joining surface.
Note that in the limit $p_y \rightarrow 0$, there is no actual $y$ ($y^{\prime}$) dependence in
$\psi^{(\mu)}_{0,0}(x,y)$ ($\psi^{(\mu)\prime}_{0,0}(x^{\prime},y^{\prime})$).
From Fig. \ref{fig:120degree_bend_join} one may see that the positions
denoted as $x(^\prime) = n(^\prime)$ demarcate increments of length $a/2$.
The meaning of the formal expression (Eq. \ref{M00}),
using Eq. \ref{arm_ribbon},
then takes the form

\begin{widetext}
\begin{eqnarray}
M_{0,0}(p_y=0) &\propto& \sum_n
\Biggl\{
\left| \exp\left[-i\frac{2\pi}{3}\left({3 \over 2} n+{1 \over 2}\right)\right]
- \exp\left[i\frac{2\pi}{3}\left({3 \over 2} n+{1 \over 2}\right)\right] \right|^2 \\
&-&\left| \exp\left[-i\frac{2\pi}{3}\left({3 \over 2} n+{1}\right)\right]
- \exp\left[i\frac{2\pi}{3}\left({3 \over 2} n+{1}\right)\right] \right|^2  \\
&+& \left |\exp\left[-i\frac{2\pi}{3}\left({3 \over 2} n+{3 \over 2}\right)\right]
- \exp\left[i\frac{2\pi}{3}\left({3 \over 2} n+{3 \over 2}\right)\right] \right|^2 =0
.
\label{overlap_back}
\end{eqnarray}
\end{widetext}

That $M_{0,0}(p_y)$ vanishes in the limit $p_y \rightarrow 0$
indicates an absence of backscattering
as $\varepsilon \rightarrow 0$, and hence perfect
transmission in this limit.  One may also compute the corresponding
overlap on the joining surface for transmission and confirm that
it has a magnitude of unity.
Deviations from this are of order
$(p_yW)^2$, so that these become significant when $p_y \sim 1/W$, which
occurs at an energy of the same order as the bottom of the first
excited subband.  Thus for energies well below this, we expect
the transmission to be very close to unity.  This behavior is confirmed
by the tight-binding calculations.

This behavior seems dramatically different than the na\"ive expectation
discussed in the Introduction, that the interchange of the {\bf K} and
${\bf K}^{\prime}$ valleys might lead one to expect a 60$^{\circ}$
deflection of the electron trajectory to be suppressed.  The mode
matching procedure however demonstrates that only the overlap
on the joining surface need be considered, and because this involves
a small subset of lattice points,
destructive interference between the rapidly oscillating
parts of the wavefunction may not be realized.

A second example of this procedure may be considered for the
geometry illustrated in Fig. \ref{fig:simple_ZZ_junction}, in which two armchair leads
are joined at the two solid lines to a short length of zigzag nanoribbon.
Traversal through the equilateral triangle with
two leads may be thought of as a special case of this
geometry, with the shortest possible zigzag ribbon.
Approximate wavefunctions
$\Phi_{p_n,p_y}^{(3)}(x_3,y_3)$
for the zigzag ribbon region may
be developed \cite{brey1}.  These are more
complicated than the armchair forms, in that
both $p_n$ and $p_y$ vary as $\varepsilon$ varies,
so that the transverse
wavefunctions vary with $\varepsilon$ even within a single
subband.  At energies close to zero, this
variation becomes quite pronounced in that the wavefunctions become highly localized
at the zigzag ribbon surfaces \cite{Fujita_1996}.

One may develop an explicit expression for the
transmission amplitude for this geometry,
within the single mode approximation, in terms of
the overlap integrals on the two junctions \cite{fertig_unpub}.
The result is proportional to the product of the overlap
integrals on each of the joining surfaces, $N^{(i)}_{0,0}(p_y,p_y^{\prime})$, with
\begin{widetext}
\begin{eqnarray}
N^{(1)}_{0,0}(p_y,p_y^{\prime})&=&
\int d\lambda_1 \psi^{(1)*}_{0,p_y}(x(\lambda_1),y(\lambda_1))
\Phi_{p_0,p_y^{\prime}}^{(3)}(x(\lambda_1),y(\lambda_1)) \\
N^{(2)}_{0,0}(p_y,p_y^{\prime})&=&
\int d\lambda_2
\Phi_{p_0,p_y^{\prime}}^{(3)*}(x(\lambda_2),y(\lambda_2))
\psi^{(2)}_{0,p_y}(x(\lambda_2),y(\lambda_2)).
\end{eqnarray}
\end{widetext}
In these integrals, $\lambda_1$ and $\lambda_2$ parameterize
the left and right surfaces in Fig. \ref{fig:simple_ZZ_junction}, $\pm p_0(\varepsilon)$
is the transverse momentum from which the lowest zigzag transverse
state is made up \cite{brey1,com2}, and
$\varepsilon^2=v_F^2p_y^2=v_F^2(p_0^2+p_y^{\prime 2})$.
The important observation in this case is that, at low energy,
the states of the zigzag ribbon $\Phi$ become confined
to the surfaces, with a length scale $\xi(\varepsilon)$
which vanishes rapidly at low energy in the continuum
description \cite{brey1}.  Since the lowest transverse
state of the armchair ribbon remains spread throughout
the ribbon cross-section, one may see $N^{(i)}_{0,0} \sim \sqrt{\xi(\varepsilon)/W}$,
and the resulting conductance will behave as $G \sim [\xi(\varepsilon)/W]^2$.
This means the conductance per spin is is suppressed near zero energy, but can
rise to a value of order $e^2/h$ once $\varepsilon$ is above
the range of energies where zigzag ribbons support surface
states.  The resulting conductance is suppressed in
a narrow range near zero energy.

As a final example, we consider the three lead equilateral
triangle geometry.  The SMA has a form identical to the
case of the zigzag ribbon described above, with the
overlap integrals now performed on surfaces joining
armchair ribbons.  Unlike the above two examples,
the surface is oriented at {\it different}
angles with respect to the cross-sections of the two
ribbons, as illustrated in Fig. \ref{fig:simple_EQT_junction}.
In this case the
two integrals $\tilde{N}^{(1,2)}_{0,0}$
whose product is proportional to the
transmission amplitude involve a product of
wavefunctions whose
fast components vary at different rates as
one moves along the joining surface.  The resulting
overlaps are very sensitive to the precise way in
which the leads are joined to the triangle.

\begin{figure}[htb]
\centering
\includegraphics[width=8.cm]{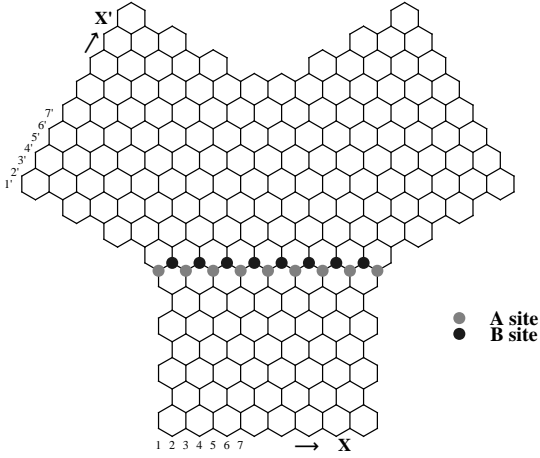}
\caption{
Diagram detailing the joining surface between the lower lead
and the upper left lead, for an equilateral triangle with three
leads, and the leads attached to the triangle at the wider
point in their cross-section.
}
\label{fig:Tri-2points}
\end{figure}

Fig. \ref{fig:Tri-2points} details a joining surface for the
geometry of Fig. \ref{fig:Triangle_3Lead_AC_2}.
Note in this case that the ribbons are joined along zigzag edges,
so that the matching of both wavefunctions and currents is
easily accomplished.
Denoting the wavefunction
in the lower lead by $\psi^{(\mu)}_{0,p_y}(x,y)$
and the one in the upper left lead by $\psi^{(\mu)\prime}_{0,p_y}(x^{\prime},y^{\prime})$,
with $\mu=A,B$, we find the matching conditions for $p_y=0$ to have the form
\begin{widetext}
\begin{equation}
\begin{array}{cc}
\psi^{(A)}_{0,0}(x=1,y)=\psi^{(A)\prime}_{0,0}(x^{\prime}=1^{\prime},y^{\prime}) & \quad,\quad
          \psi^{(B)}_{0,0}(x=2,y)=\psi^{(B)\prime}_{0,0}(x^{\prime}=2^{\prime},y^{\prime}) \\
\psi^{(A)}_{0,0}(x=3,y)=\psi^{(A)\prime}_{0,0}(x^{\prime}=2^{\prime},y^{\prime}) & \quad,\quad
          \psi^{(B)}_{0,0}(x=4,y)=\psi^{(B)\prime}_{0,0}(x^{\prime}=3^{\prime},y^{\prime}) \\
\psi^{(A)}_{0,0}(x=5,y)=\psi^{(A)\prime}_{0,0}(x^{\prime}=3^{\prime},y^{\prime}) & \quad,\quad
          \psi^{(B)}_{0,0}(x=6,y)=\psi^{(B)\prime}_{0,0}(x^{\prime}=4^{\prime},y^{\prime}) \\
\cdot & \quad\quad \cdot\\
\cdot & \quad\quad \cdot\\
\cdot & \quad\quad \cdot\\
\end{array}
\label{match-2points}
\end{equation}
\end{widetext}
Again, for $p_y=0$ there is no actual $y$ ($y^{\prime}$) dependence in
$\psi^{(\mu)}_{0,0}(x,y)$ ($\psi^{(\mu)\prime}_{0,0}(x^{\prime},y^{\prime})$).
The overlap between the incoming state from the bottom lead
and the outgoing state from the upper left lead, using Eq. \ref{arm_ribbon},
has the form
\begin{widetext}
\begin{eqnarray}
\tilde{N}^{(1)}_{0,0} &\sim& {1 \over W} \sum_n
\Biggl\{
\left[ \exp\left(-i\frac{2\pi n}{3}\right) - \exp\left(i\frac{2\pi n}{3}\right) \right]
\left[
\exp\left(i\frac{4\pi}{3}(n+{1 \over 2})\right) -
         \exp\left(-i\frac{4\pi}{3}(n+{1 \over 2})\right)
         \right]
         \nonumber\\
         &-& \left[ \exp\left(-i\frac{2\pi}{3}(n+1)\right) -
                    \exp\left(i\frac{2\pi}{3}(n+1)\right)
\right]
             \left[\exp\left(i\frac{4\pi}{3}n\right)-\exp\left(-i\frac{4\pi}{3}n\right)\right]
\Biggr\}.
\label{overlap1}
\end{eqnarray}
\end{widetext}
The upper line in Eq. \ref{overlap1} is due to the overlap of the wavefunctions
on the $A$ sites, while the lower line comes from the $B$ sites.
Multiplying out the square brackets generates terms which are either
independent of the integer $n$ or oscillate in $n$ with period 3.
One finds that the non-oscillating terms from the upper and lower lines
precisely cancel.  The remaining rapidly oscillating terms vanish in the sum
provided the maximum value of $n$ is a multiple
of 3, and in any case give vanishing contribution as the ribbons become wide ($W \rightarrow \infty$).
The cancellation of the non-oscillating terms
indicates a complete destructive interference between incoming and
outgoing waves for the two arms of the triangle at low energy, as one might have na\"ively
supposed.  With no overlap at the joining surface,
the conductance should vanish at zero energy.

\begin{figure}[htb]
\centering
\includegraphics[width=8.cm]{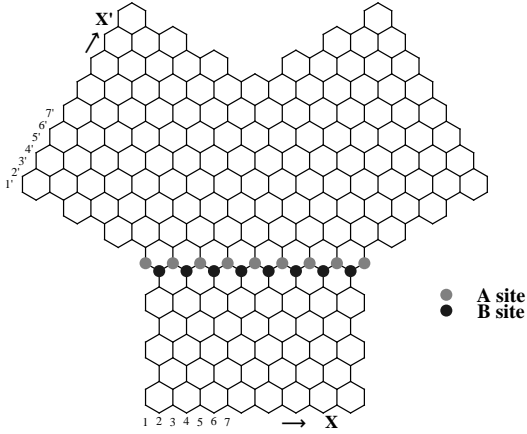}
\caption{
Diagram detailing the joining surface between the lower lead
and the upper left lead, for an equilateral triangle with three
leads, and the leads attached to the triangle at the narrower
point in their cross-section.
}
\label{fig:Tri-1point}
\end{figure}

The other simple geometry for joining the ribbons to the triangle is detailed
in Fig. \ref{fig:Tri-1point}.  In this case the matching conditions take
the form
\begin{widetext}
\begin{equation}
\begin{array}{cc}
\psi^{(A)}_{0,0}(x=1,y)=\psi^{(A)\prime}_{0,0}(x^{\prime}=1^{\prime},y^{\prime}) & \quad,\quad
          \psi^{(B)}_{0,0}(x=2,y)=\psi^{(B)\prime}_{0,0}(x^{\prime}=1^{\prime},y^{\prime}) \\
\psi^{(A)}_{0,0}(x=3,y)=\psi^{(A)\prime}_{0,0}(x^{\prime}=2^{\prime},y^{\prime}) & \quad,\quad
          \psi^{(B)}_{0,0}(x=4,y)=\psi^{(B)\prime}_{0,0}(x^{\prime}=2^{\prime},y^{\prime}) \\
\psi^{(A)}_{0,0}(x=5,y)=\psi^{(A)\prime}_{0,0}(x^{\prime}=3^{\prime},y^{\prime}) & \quad,\quad
          \psi^{(B)}_{0,0}(x=6,y)=\psi^{(B)\prime}_{0,0}(x^{\prime}=3^{\prime},y^{\prime}) \\
\cdot & \quad\quad \cdot\\
\cdot & \quad\quad \cdot\\
\cdot & \quad\quad \cdot\\
\end{array}
\label{match-1point}
\end{equation}
\end{widetext}
and the corresponding overlap sum is now

\begin{widetext}
\vbox{
\begin{eqnarray}
\tilde{N}^{(1)}_{0,0} &\sim& {1 \over W} \sum_n
\Biggl\{
\left[ \exp\left(-i\frac{2\pi n}{3}\right) - \exp\left(i\frac{2\pi n}{3}\right) \right]
\left[
\exp\left(i\frac{4\pi}{3}(n+{1 \over 2})\right) -
         \exp\left(-i\frac{4\pi}{3}(n+{1 \over 2})\right)
         \right]
         \nonumber\\
         &-& \left[ \exp\left(-i\frac{2\pi}{3}n\right) -
                    \exp\left(i\frac{2\pi}{3}n\right)
\right]
             \left[\exp\left(i\frac{4\pi}{3}n\right)-\exp\left(-i\frac{4\pi}{3}n\right)\right]
\Biggr\}\nonumber\\
&=&{1 \over W}\sum_n\left[-2\cos\left({{2\pi} \over 3}\right)+2 + ({\rm ~oscillating ~terms}) \right]
.
\label{overlap2}
\end{eqnarray}
}
\end{widetext}

We see the slight shift in positions of the $B$ sites where the wavefunctions
are matched in Eq. \ref{match-1point} leads to an extra phase factor, such
that the $A$ and $B$ overlaps no longer cancel.  Thus $\tilde{N}^{(1)}_{0,0}$
is relatively large in this case, and we expect a correspondingly large
transmission.

We next turn to our numerical studies, which, as discussed in the
Introduction, essentially confirm the expectations of the SMA.

\section{Model and Numerical Method}

\begin{figure}[htb]
\centering
\includegraphics[width=6 cm]{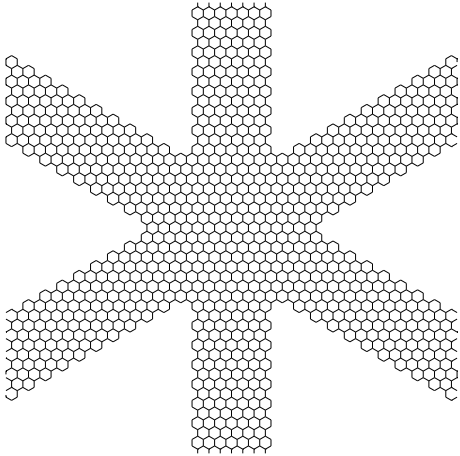}
\caption{Hexagon with 6 leads
}
\label{fig:Hex_6Lead}
\end{figure}

Our calculations are based on a simple tight-binding model of graphene,
in which only nearest neighbor hopping is included.  Formally
the Hamiltonian may be written as
\begin{equation}
H= - t \sum_{\{ {\bf R}_i{\bf R}_j \}} \left(
\ket{{\bf R}_i} \bra{{\bf R}_j} +
\ket{{\bf R}_j} \bra{{\bf R}_i}
\right).
\label{hamiltonian}
\end{equation}
Here $t$ is the nearest-neighbor hopping matrix element and
$\{ {\bf R}_i{\bf R}_j \}$ denotes bonds on a honeycomb
lattice with boundaries defined by the polygon of interest
and/or the attached leads.
For the
purposes of this section only we use the energy unit $t = 1$.  
An example of a hexagonal
scattering region with attached leads
is illustrated in Fig. \ref{fig:Hex_6Lead}.
We consider only armchair and zigzag nanoribbons for our leads, although
other periodic edges may be considered within our method.  We wish to calculate
conductances through various pairs of leads in this geometry.  Below
we discuss the procedures used to evaluate the conductance
per spin
$G$ from Green's functions
of the Hamiltonian $H$, under the assumption of time-reversal symmetry.

\begin{figure}[t]
\centering
\includegraphics[width=\columnwidth]{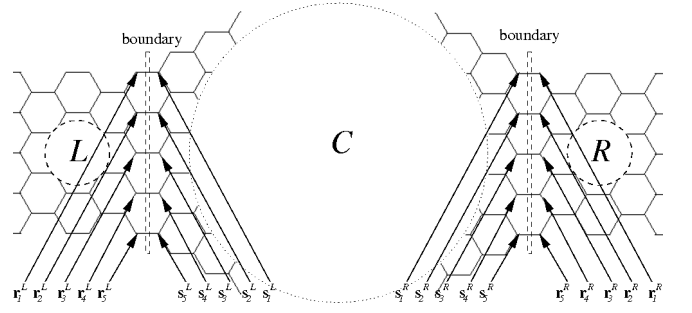}
\caption{Conceptual division of the geometry into regions
$C$, $L$, and $R$, illustrating the lattice point labeling scheme 
at the boundaries, which are denoted by pairs of dashed lines. This example is a
symmetric armchair ribbon with $N=4$, and the boundary is chosen
such that $Q=5$.}
\label{fig:CLR}
\end{figure}

Conceptually, we divide the lattice into three regions: a central region $C$,
a ``left'' lead $L$ and a ``right'' lead $R$, as illustrated in
Fig. \ref{fig:CLR} for the case of armchair nanoribbon leads.
We compute the conductance between the leads  $L$ and $R$,
and any remaining leads in the geometry are considered  part of the
central region $C$.

Conductance at zero temperature is computed as a linear response using
the Kubo formula \cite{mahan},
which relates conductivity to a current-current
correlation function.  For the two-lead conduction problem, the
relevant current operators correspond physically to the
total current in the region $L$ flowing in the incoming direction
and the total current in the region $R$ flowing in the outgoing
direction. We denote these operators as
$\hat{J}_L$ and $\hat{J}_R$, whose precise definition we elucidate
below.  Charge conservation implies that
the current flux down a given lead in a time-independent, steady state
is the same regardless of
where in the lead it is ``measured''.
For our purposes, it is most convenient to ``measure'' the current
precisely at the two boundaries where the
the respective leads join the central region. These boundaries
are each illustrated in Fig. \ref{fig:CLR} as a pair of dashed lines.

These considerations motivate the following definitions:
we label the lattice points immediately to either side of this boundary
as ${\bf r}_{\mu}^{L(R)}$  for lattice points belonging to region $L(R)$
and ${\bf s}_{\mu}^{L(R)}$ for those belonging to the central region. Here
$\mu$ runs over integers 1 to $Q$, where $Q$ is the number of bonds
traversing the boundary. Typically $Q$ is the same as the number of channels 
$N$, but $Q > N$ is possible in
certain configurations. This labeling scheme is illustrated in
Fig. \ref{fig:CLR}.  The current operators may be explicitly
defined as

\begin{equation}
\hat{J}_{L(R)} \equiv +i (-i) \mathop{\sum_{\mu = 1}^{Q}}
\left(
\ket{{\bf r}_{\mu}^{L(R)}}\bra{{\bf s}_{\mu}^{L(R)}}
-
\ket{{\bf s}_{\mu}^{L(R)}}\bra{{\bf r}_{\mu}^{L(R)}}
\right)
\end{equation}

The Kubo formula for conductivity leads to the well-known
\cite{fisher} transmission formula for the conductance per spin
\begin{equation}
G = \frac{e^2}{2 h}{\rm Tr}( \hat{t}_{LR}^{\dagger} \hat{t}_{LR} + \hat{t}_{RL}^{\dagger} \hat{t}_{RL})
\label{eq:Lee}
\end{equation}
where $\hat{t}_{LR(RL)}$ is the transmission amplitude for states
at the Fermi energy $E_F$ from the region
$L$ to $R$ ($R$ to $L$) and $\hat{t}_{LR} = \hat{t}_{RL}^{\dagger}$.
In our context, the Kubo formula involves the operators $\hat{J}_L$ and
$\hat{J}_R$, and the resulting transmission amplitudes in (\ref{eq:Lee})
are $Q \times Q$ matrices.

To evaluate the transmission amplitude, the relevant
retarded Green's functions are $Q \times Q$ matrices whose
components are given by
\begin{eqnarray}
\G[\alpha,\beta]_{\mu \nu} &\equiv&
\bra{{\bf s}_{\mu}^{\alpha}}
(E_F + i \eta - H)^{-1}
\ket{{\bf s}_{\nu}^{\beta}} \\
\G^C[\alpha,\beta]_{\mu \nu} &\equiv&
\bra{{\bf s}_{\mu}^{\alpha}}
(E_F + i \eta - H^C)^{-1}
\ket{{\bf s}_{\nu}^{\beta}} \\
\G^L_{\mu \nu} &\equiv&
\bra{{\bf r}_{\mu}^L}
(E_F + i \eta - H^L)^{-1}
\ket{{\bf r}_{\nu}^L} \\
\G^R_{\mu \nu} &\equiv&
\bra{{\bf r}_{\mu}^R}
(E_F + i \eta - H^R)^{-1}
\ket{{\bf r}_{\nu}^R} .
\end{eqnarray}
Here $H^L$, $H^R$, $H^C$ represent the restriction of the
Hamiltonian to regions $L$, $R$, and $C$, respectively (with
no hopping across the boundaries), and $\alpha$ and $\beta$
can be $L$ or $R$. Ideally, $\eta \rightarrow 0^+$, but
for numerical calculations it is necessary to choose a
small $\eta > 0$ which in effect becomes the
energy resolution of the computation. Our calculations are
carried out with $\eta = 10^{-6} t$.

With these definitions, we have, in abbreviated notation for the
$Q\times Q$ matrices,
\begin{eqnarray}
\hat{t}_{LR} &=& \hat{v}_{R}^{1/2}\hat{\G}[R,L]\hat{v}_{L}^{1/2} \\
\hat{v}_{L(R)} &\equiv& i (\hat{\G}^{L(R)} - \hat{\G}^{L(R)\dagger}).
\end{eqnarray}
Schematically, this formula shows that the transmission amplitude is
given by the propagator from the left to right side of the central region,
with the $\hat{v}_{L(R)}$ velocity matrices
\endnote{The matrices $\hat{v}_{L(R)}$ are Hermitian and
positive semi-definite.  Thus, one may uniquely construct
the ``square-root'' $\hat{v}^{1/2}_{L(R)}$ with the same
eigenvectors, but whose corresponding eigenvalues
are the positive square root.}
normalizing the nanoribbon states to unit flux.
The problem is now reduced to computing $\hat{\G}^{L(R)}$
and $\hat{\G}^C$.  Standard gluing formulas \cite{Sols} for non-interacting
Green's functions can be used to obtain $\hat{\G}[R,L]$ from these two.

For our geometries, $\hat{\G}^{L(R)}$ is the Green's function of
a semi-infinite nanoribbon at its termination.
One may directly evaluate the Green's function of a single nanoribbon
unit cell by inverting the matrix $(E_F + i\eta - h)$, where
$h$ is its lattice Hamiltonian.
We then rapidly extend from the unit cell to a
nanoribbon segment of length $2^l$ through $l$ successive
length-doubling steps via the aforementioned gluing formulas. We find that the
Green's function on one termination of the long nanoribbon segment becomes an accurate
approximation for the semi-infinite ribbon when $2^l \sim t/\eta$.
This can be verified by substituting the numerical result into a Dyson equation
satisfied by the exact Green's function. Such a Dyson equation may be
easily derived for any such semi-infinite periodic structure.

For $\hat{\G}^C$ we first suppose that the geometry has only two leads
and thus the region $C$ is finite.
One approach is to perform the matrix inversion of $(E_F + i \eta - H^C)$.
However, significant computational savings are possible when the Green's function
is only required at the boundary \cite{Iyengar_unpublished}.
In the case of more than two leads, $\G^C$ can be found by gluing
the Green's functions for the
``passive'' leads in the calculation to the Green's function of the
finite scattering region. This procedure correctly accounts for
current which is drained away by the extra leads.

\begin{figure}[htb]
\centering
\includegraphics[width=8 cm]{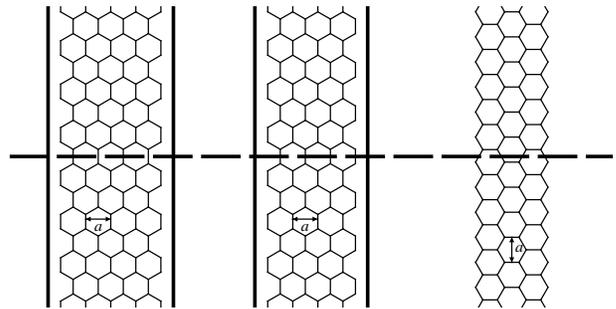}
\caption{
Symmetric armchair, asymmetric armchair, and zigzag ribbons with $N=4$.
Dashed line cuts through four bonds of each ribbon, the minimal number
required to sever each into two disconnected pieces. Heavy vertical lines
indicate rows of fictive lattice points along which the wavefunction vanishes
for the armchair nanoribbons.
}
\label{fig:ribbon_break}
\end{figure}

We first tested our numerical techniques on the straightforward case of
infinite nanoribbons, whose bandstructure is well-understood \cite{brey1}.
Nanoribbons may be characterized by the number $N$ of conducting channels.
The integer $N$ gives two related properties of the ribbon:
1) the minimum number of severed bonds required to break the ribbon into
two disconnected pieces (as in Fig. \ref{fig:ribbon_break}) and
2) the maximum possible value of $G$, the conductance per spin, in units of the
conductance quantum $e^2/h$.
Armchair nanoribbons have associated with them two lines of fictive
lattice points (just outside the actual ribbon edges)
separated by a width $L a$ (see Fig. \ref{fig:ribbon_break})
on which the wavefunction vanishes.
When the ribbons
possess a reflection symmetry through the center, $N = L - 1$, and
otherwise $N = L - 1/2$.
In either case, the ribbon is semiconducting (i.e. there is gap
in the spectrum around zero energy) unless $\sin(4\pi L/3)=0$.
Thus, symmetric armchair nanoribbons are metallic only for the series
$N=2,5,8,\ldots$ and the asymmetric ribbons for $N=1,4,7,10,\ldots$.
Zigzag ribbons are metallic for any value of $N$.

We have compared the conductance of these various types of ribbons
as a function of $E_F$ with their bandstructure. We find, as expected,
a contribution to $G(E_F)$ of $e^2/h$ for each band present at
$E_F$, with the exception of the flat bands\cite{flatbands}
in symmetric armchair ribbons
at energies $\pm t$. We also verify the linear dispersion of the
metallic band in armchair nanoribbons with velocity
$v_F = \sqrt{3} t a /2$, as well as the maximum $G$ values and
metallic $N$ sequences discussed above.


\section{Numerical Results}

\subsection{Armchair Lead Systems: $120^{\circ}$ Junctions and Triangles}

Our low-energy result for the simple
junction for armchair leads at low energy was discussed in the
Introduction, and is illustrated in Fig. \ref{fig:120Bend_AC}.
One clearly sees that the tranmission is essentially
perfect, with a small suppression just below the first
excited subband energy. The behavior appears to be well-explained by the
SMA.  The $120^{\circ}$ junction also naturally occurs 
in an equilateral triangle with two leads,
as illustrated in Fig. \ref{fig:Triangle_2Lead_AC_Diagram}.
As has already been explained, at low energies one finds
suppression of the conductance in a narrow range
around zero energy.

We discuss more fully the example of the three lead triangle
geometry,
illustrated in Fig. \ref{fig:Triangle_3Lead_AC_2}(d).
Because of the symmetry of this geometry, the two point conductance
is the same for any pair of the three leads.
Fig. \ref{fig:Triangle_FullBand}
illustrates
the conductance for three different system sizes as a function
of Fermi energy $E_F$ over the entire bandwidth.
One may see that the overall shape of the conductance
curve roughly tracks the density of states for bulk graphene,
with peaks at the van Hove singularities $E_F=t$, where
$t$ is the hopping matrix element. The bumps
and wiggles around this are due to changes in the number
of conducting channels as the Fermi energy is increased, as
well as to quantum interference in the scattering region \cite{datta}.

\begin{psfrags}
\psfrag{xlabel}[][][0.7]{$E_F/t$}
\psfrag{ylabel}[b][cl][0.7]{G $(e^2/h)$}
\begin{figure}[htb]
\centering
\subfigure[$N$=5]{
\includegraphics[width=\figwidth]{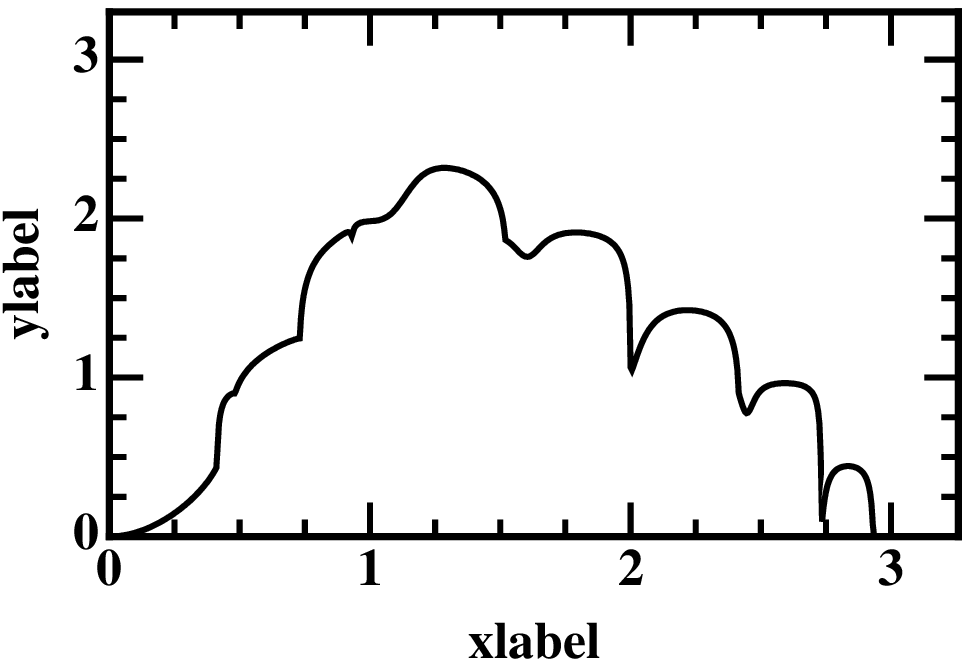}
}
\subfigure[$N$=10]{
\includegraphics[width=\figwidth]{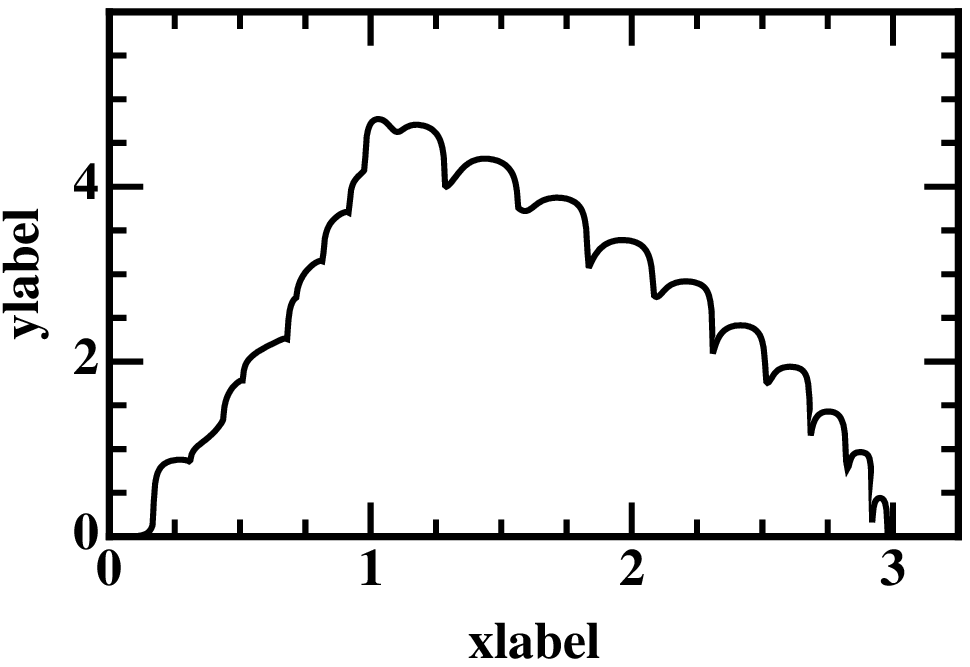}
}\\
\subfigure[$N$=15]{
\includegraphics[width=\figwidth]{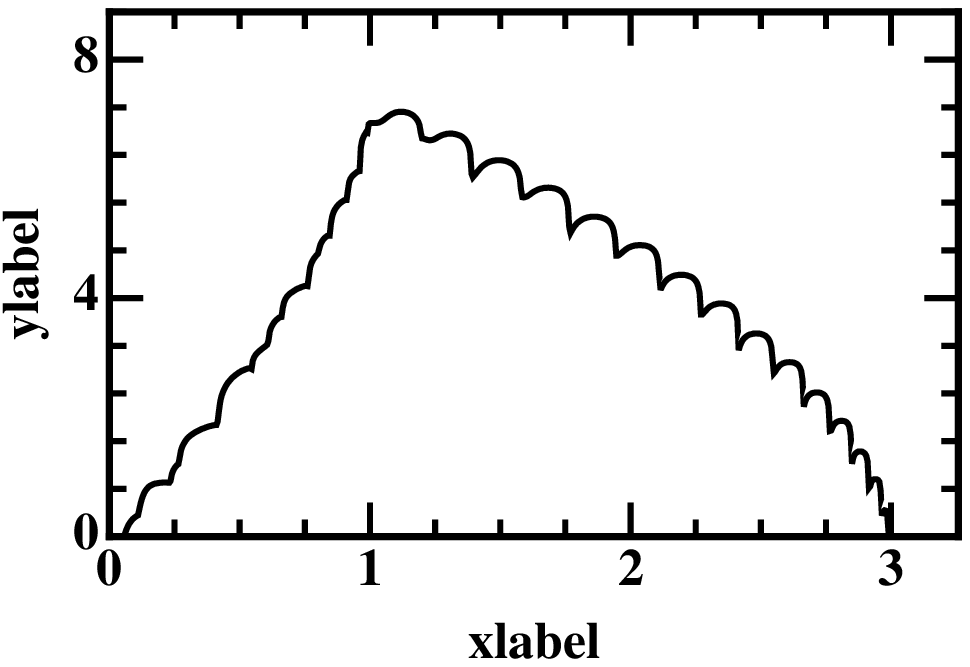}
}
\caption{Conductance per spin through an
equilateral triangle with three armchair leads of sizes
(a) $N$=5, (b) $N$=10, and (c) $N$=15.
}
\label{fig:Triangle_FullBand}
\end{figure}
\end{psfrags}

The low energy region of conductance for different system sizes may
also be considered.
In Fig. \ref{fig:Triangle_3Lead_AC_lowenergy}
we blow up the low-energy
region for three different system sizes.
The cases $N=13$ and $N=15$ as expected reveal no
conduction at low energies, since there are no conducting states
to carry current through the leads.  The metallic state for $N=14$,
by contrast, allows a finite conductance, but one may see its
actual value is remarkably small at very low
energy.  One may examine this behavior with increasing
$N$, and not surprisingly the pattern of near suppression
for system sizes of the form $N=3M+2$, and complete suppression
for other sizes, repeats itself.  When viewed as a function of
$E_F/E_e$, which fixes the position of the opening of the
first excited subband as $N$ becomes large, the conductance in
the lowest subband tends to a roughly parabolic shape, very small
but remaining finite away from $E_F=0$ as $N \rightarrow \infty$.
This represents the continuum limit, in which the width $W$ of the
ribbons remain finite, and the lattice constant $a$ is taken to
zero.  The increase from zero of the conductance as the energy
increases from zero is in agreement with the results of the SMA,
but we note that the vanishing conductance for these widths
requires the joining geometry illustrated in Fig. \ref{fig:Tri-2points}.
As discussed in the Introduction, a joining geometry of the
form illustrated in Fig. \ref{fig:Tri-1point} leads to a non-vanishing
conductance (Fig. \ref{fig:Triangle_3Lead_AC_1}),
in agreement with the SMA analysis.

\begin{psfrags}
\psfrag{xlabel}[][][0.7]{$E_F/E_e$}
\psfrag{ylabel}[b][cl][0.7]{G $(e^2/h)$}
\begin{figure}[htb]
\centering
\subfigure[$N$=13]{
\includegraphics[height=2.5cm]{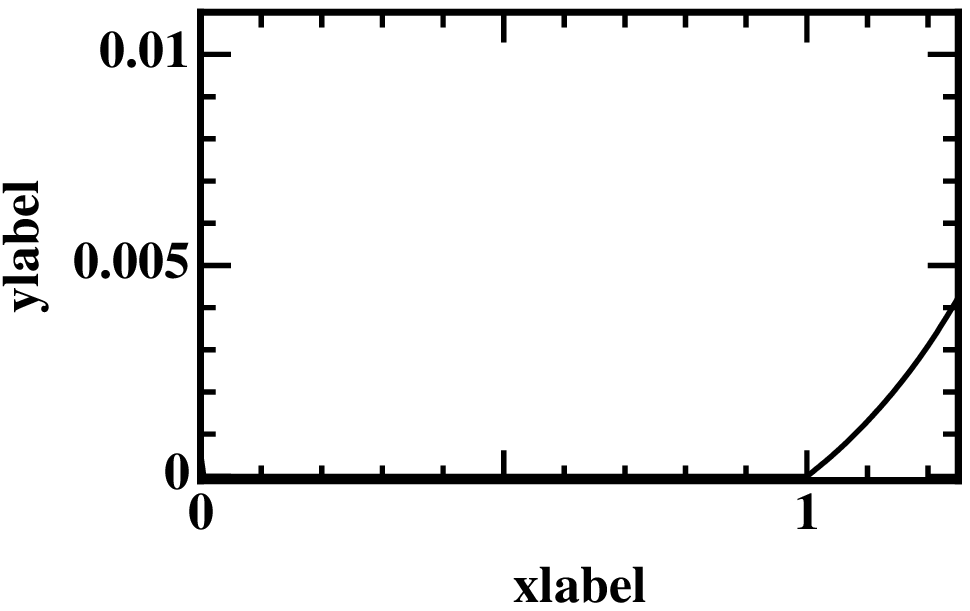}
}
\subfigure[$N$=14]{
\includegraphics[height=2.5cm]{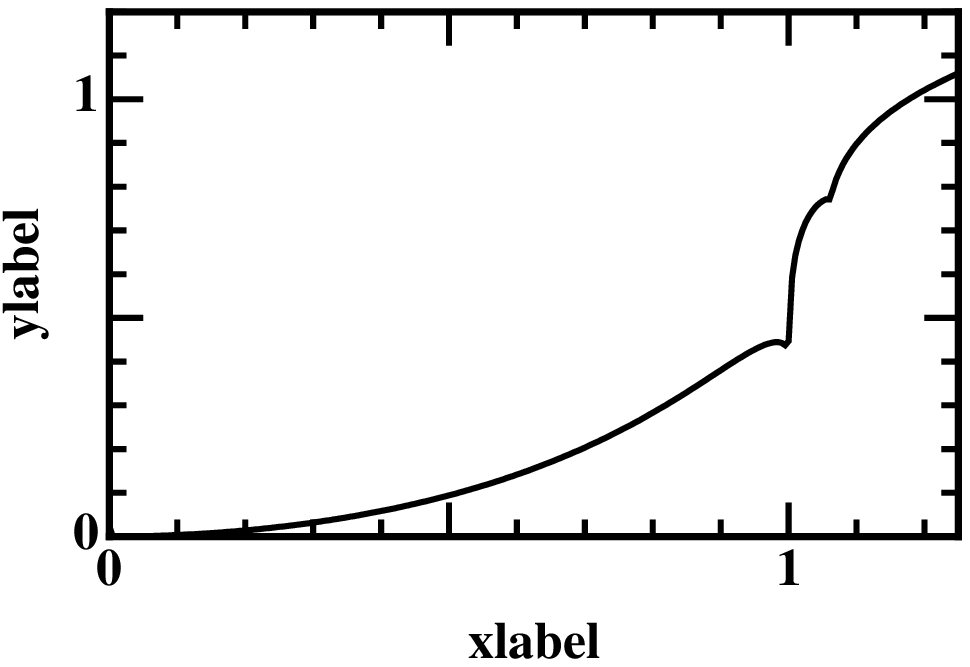}
}\\
\subfigure[$N$=15]{
\includegraphics[height=2.5cm]{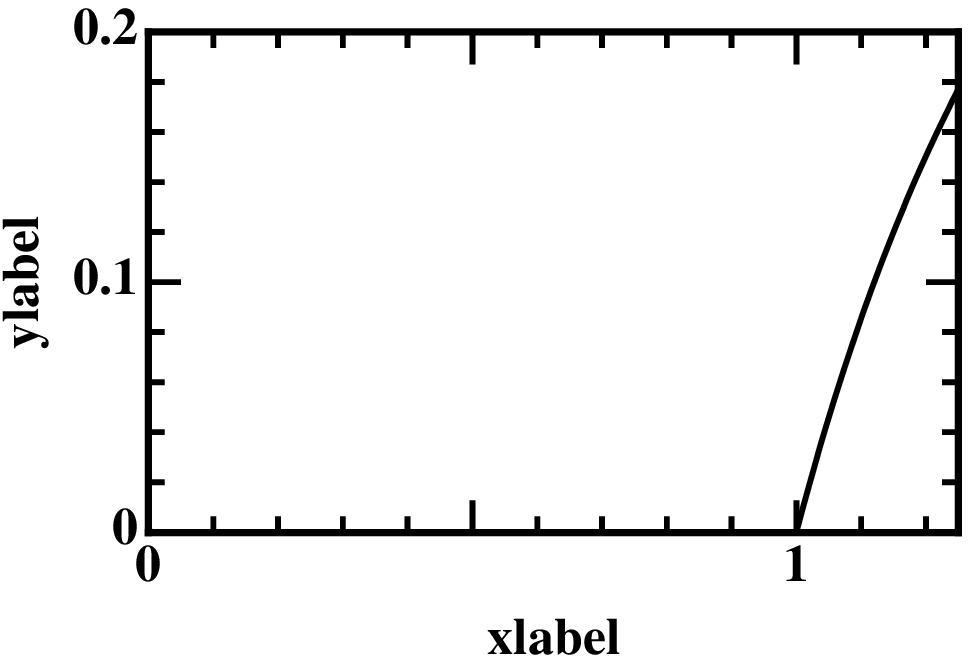}
}
\caption{ Conductance per spin in the low energy region for an
equilateral triangle with three armchair leads. (a) $N$=13, (b) $N$=14, (c) $N$=15.
}\label{fig:Triangle_3Lead_AC_lowenergy}
\end{figure}
\end{psfrags}

\subsection{Transmission Through a Hexagon}

We next consider the case of a hexagon with six attached leads,
as illustrated schematically in Fig. \ref{fig:Hex_6Lead}.
The conductance per spin is illustrated for the full
bandwidth in Figs. \ref{fig:Hex_FullBand_60}, \ref{fig:Hex_FullBand_120}, and
\ref{fig:Hex_FullBand_180}. Here we must
specify the angle between the two leads upon which
the measurement is made: Fig. \ref{fig:Hex_FullBand_60} corresponds to a
60$^{\circ}$ angle between leads, Fig. \ref{fig:Hex_FullBand_120} to 120$^{\circ}$,
and Fig. \ref{fig:Hex_FullBand_180} to 180$^{\circ}$.
Several remarks are in order.  As in the equilateral triangle,
the overall structure of the conductance follows the
density of states for bulk graphene.  However, as the size
of the system increases it is apparent that there is a
remarkably sharp suppression for transmission
through 60$^{\circ}$ and 180$^{\circ}$, and a strong
enhancement for transmission through 120$^{\circ}$,
at the van Hove singularity, $E_F/t=1$.  The rapid change
in resistance with respect to Fermi energy suggests
that this phenomenon could in principle be useful
as a transistor, although the relatively high
Fermi energy where it occurs may require a large
electric field to realize.  Beyond this, it is also
noteworthy that the overall scale of transmission through
120$^{\circ}$ is significantly larger, and seems to grow more
quickly with system size, than for the other two directions.


\begin{psfrags}
\psfrag{xlabel}[][][0.7]{$E_F/t$}
\psfrag{ylabel}[b][cl][0.7]{G $(e^2/h)$}
\begin{figure}[htb]
\centering
\subfigure[$N$=5]{
\includegraphics[width=\figwidth]{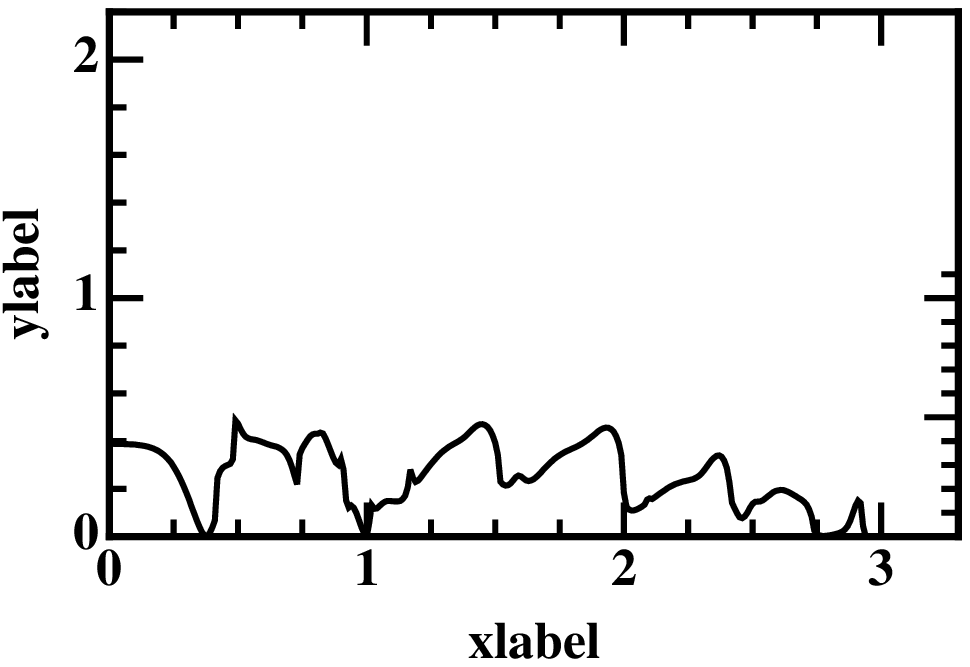}
}
\subfigure[$N$=15]{
\includegraphics[width=\figwidth]{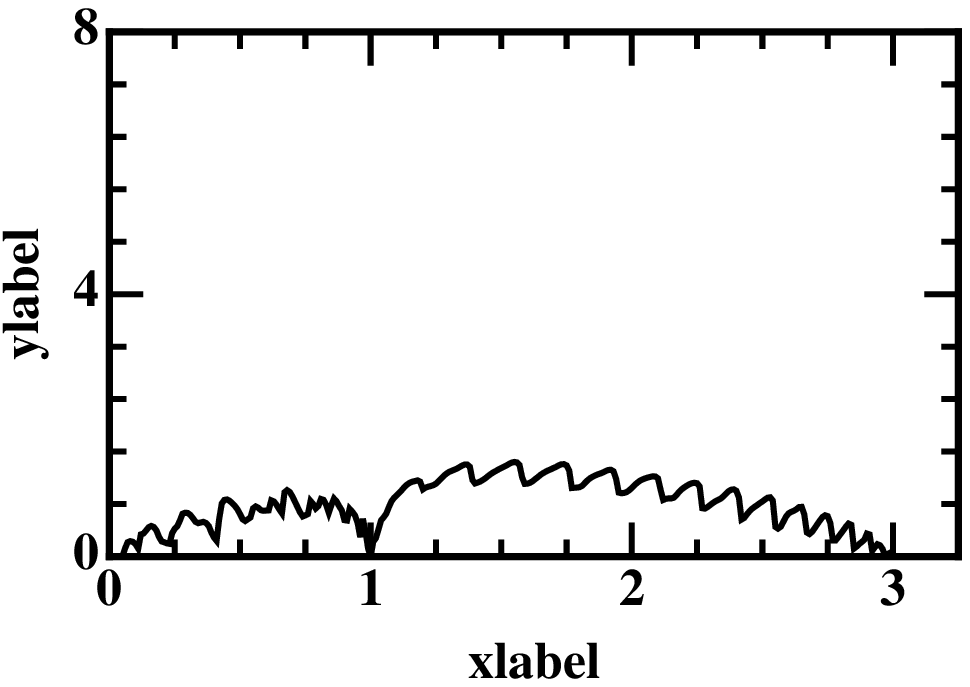}
}\\
\subfigure[$N$=20]{
\includegraphics[width=\figwidth]{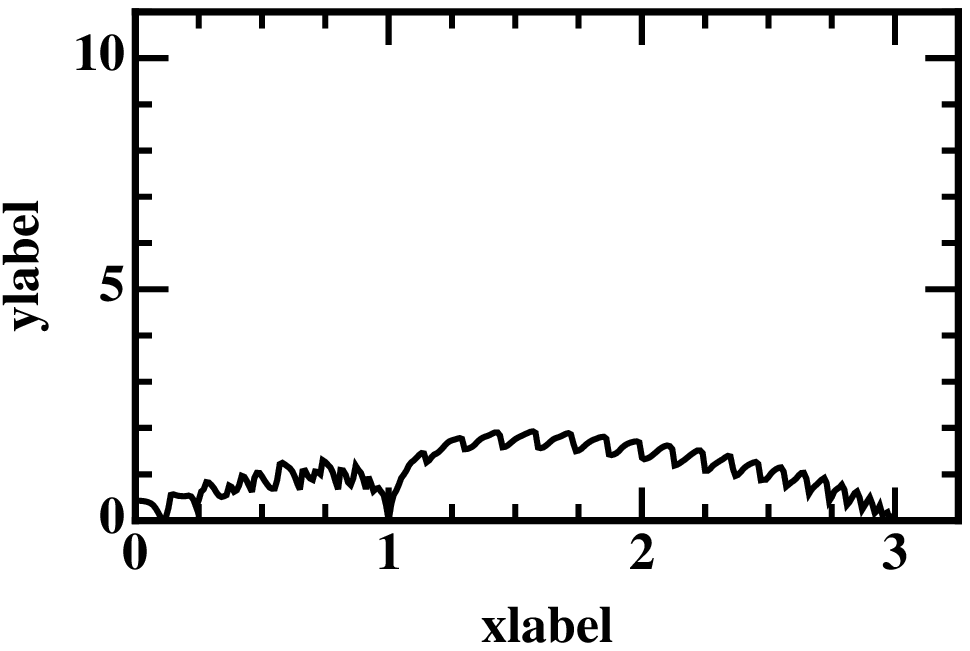}
}
\caption{Conductance per spin for two leads at an angle of 60$^{\circ}$, through a
hexagon with six armchair leads.  (a) $N$=5, (b) $N$=15, (c) $N$=20.
}
\label{fig:Hex_FullBand_60}
\end{figure}

\begin{figure}[htb]
\centering
\subfigure[$N$=5]{
\includegraphics[width=\figwidth]{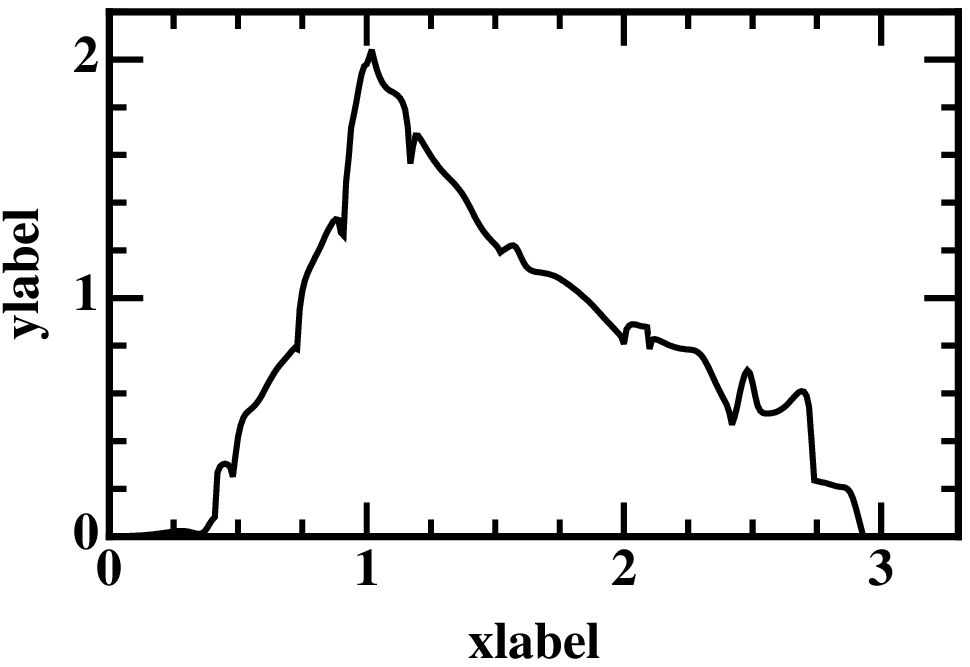}
}
\subfigure[$N$=15]{
\includegraphics[width=\figwidth]{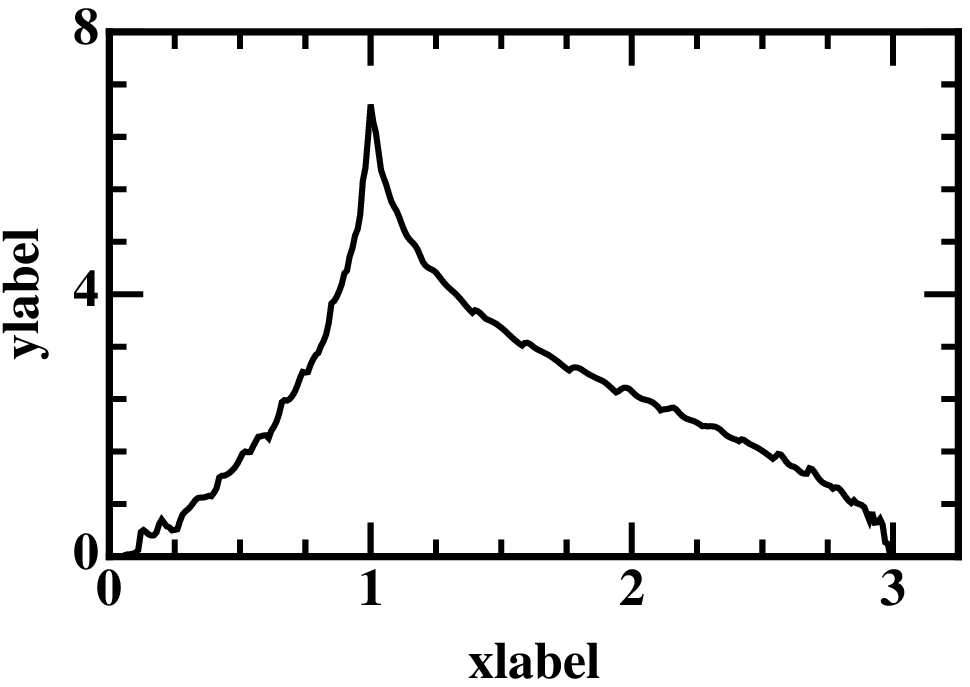}
}\\
\subfigure[$N$=20]{
\includegraphics[width=\figwidth]{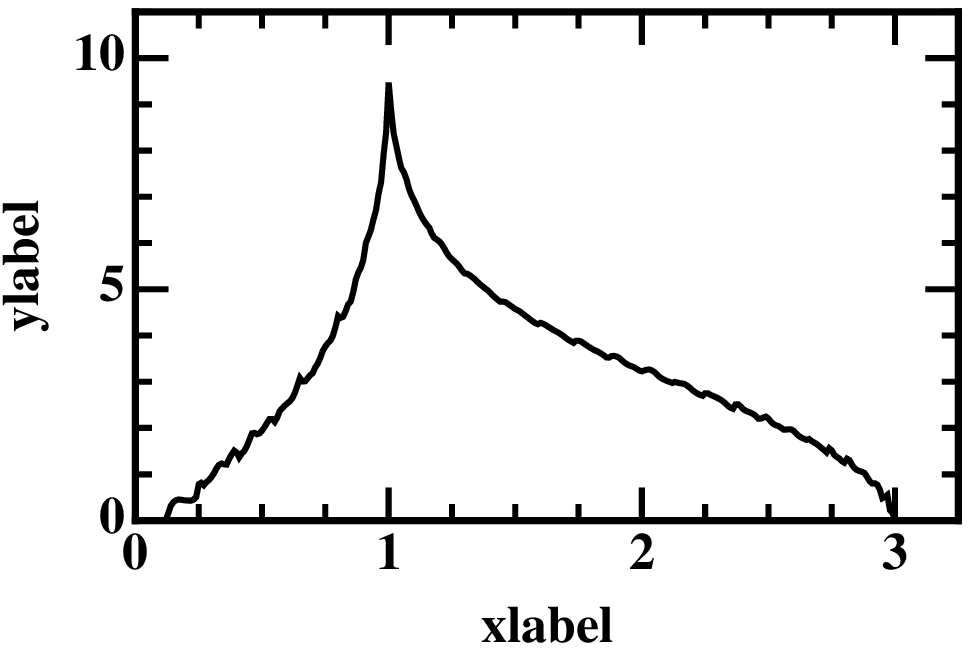}
}
\caption{
Conductance per spin for two leads at an angle of 120$^{\circ}$, through a
hexagon with six armchair leads.  (a) $N$=5, (b) $N$=15, (c) $N$=20.
}
\label{fig:Hex_FullBand_120}
\end{figure}

\begin{figure}[htb]
\centering
\subfigure[$N$=5]{
\includegraphics[width=\figwidth]{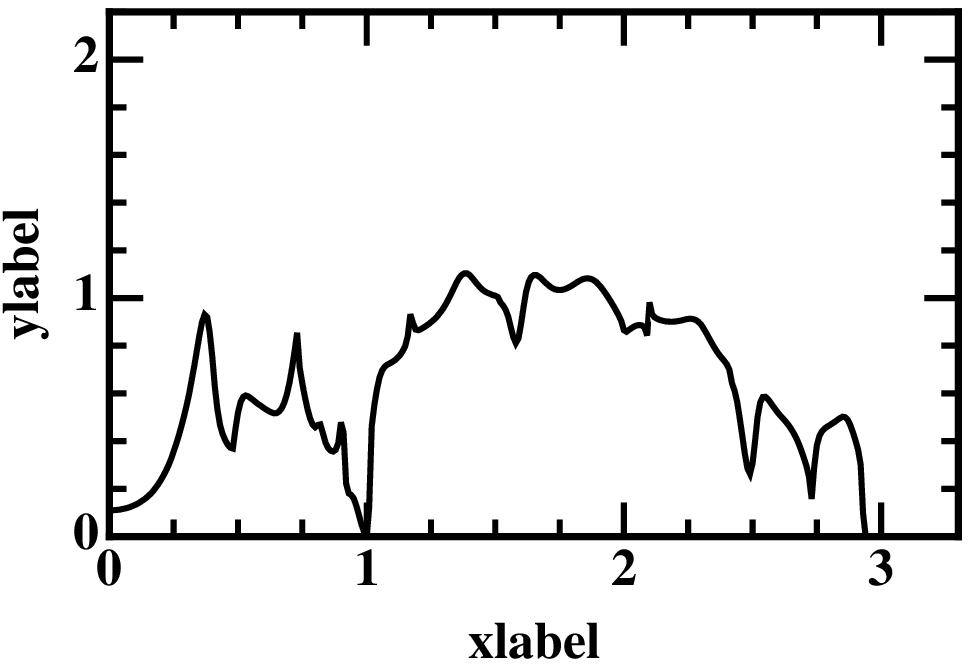}
}
\subfigure[$N$=15]{
\includegraphics[width=\figwidth]{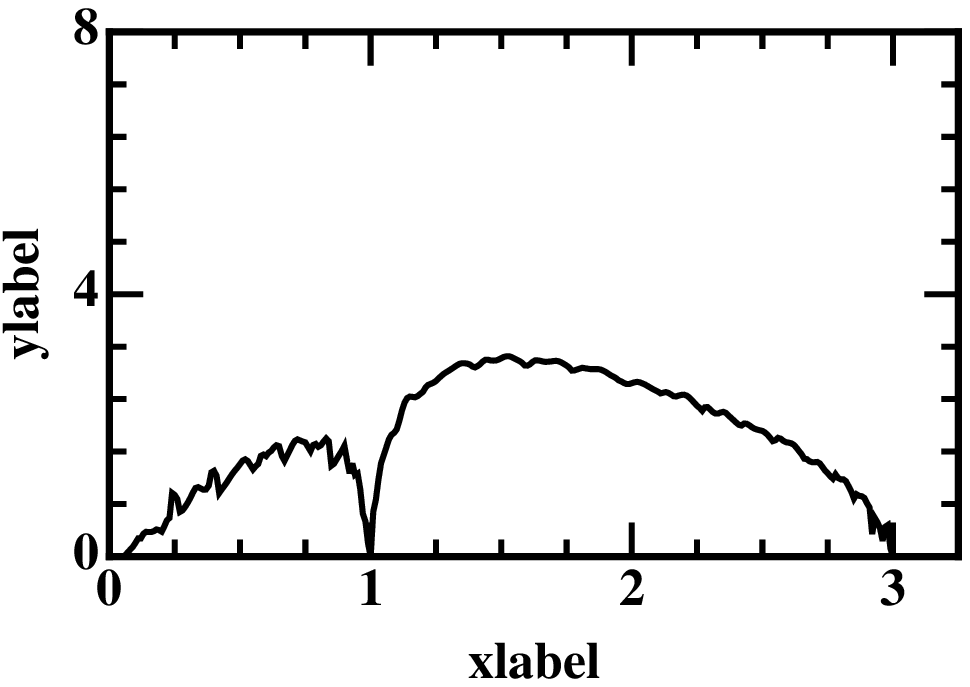}
}\\
\subfigure[$N$=20]{
\includegraphics[width=\figwidth]{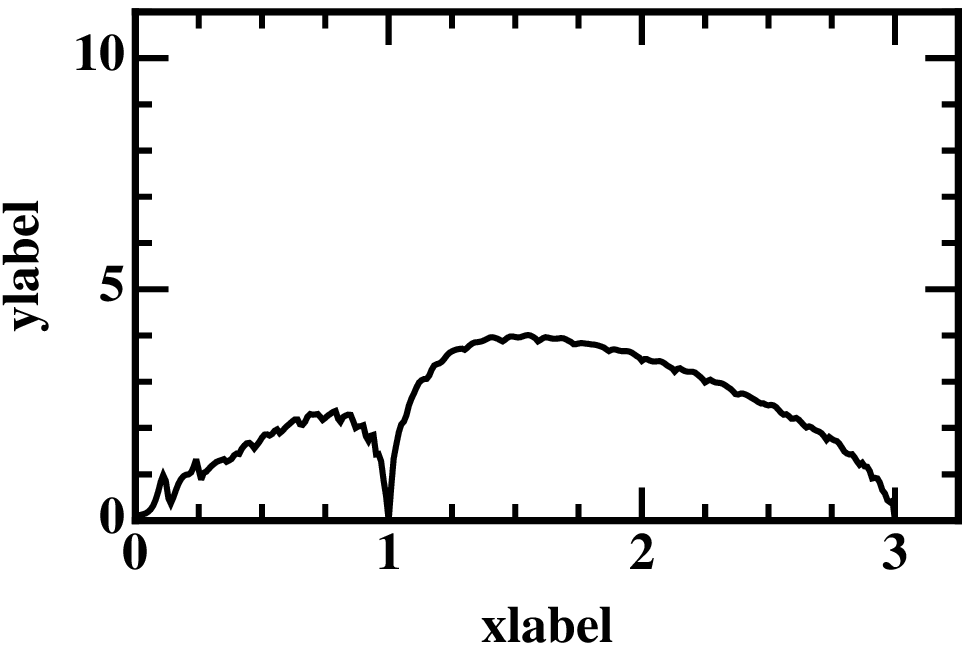}
}
\caption{
Conductance per spin for two leads at an angle of 180$^{\circ}$, through a
hexagon with six armchair leads.  (a) $N$=5, (b) $N$=15, (c) $N$=20.
}
\label{fig:Hex_FullBand_180}
\end{figure}
\end{psfrags}

The conductance at low-energy for this geometry is illustrated
in Fig. \ref{fig:Hex_AC_N=20_lowenergy} for $N=20$, for conductance through each of the
three angles.
A
suppression of 120$^{\circ}$ transmission is apparent, and in this
case is in fact much more pronounced throughout the first
subband than is the case for the equilateral triangle.
Indeed, in studies of hexagons with only three or two leads,
we find the 120$^{\circ}$ transmission to be even more suppressed
throughout the lowest subband than in the six lead case.
This suggests the hexagon may be useful in three terminal
devices where one may wish to employ one lead as a voltage
probe without draining current flowing between other leads.
\begin{psfrags}
\psfrag{xlabel}[][][0.7]{$E_F/E_e$}
\psfrag{ylabel}[b][cl][0.7]{G $(e^2/h)$}
\begin{figure}[htb]
\centering
\subfigure[60$^{\circ}$]{
\includegraphics[width=\figwidth]{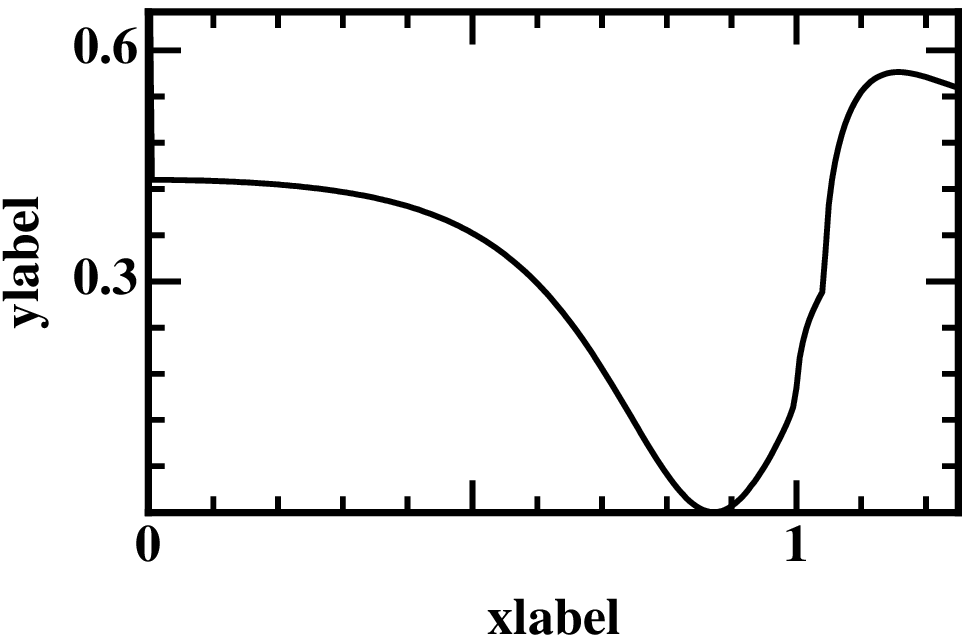}
}
\subfigure[120$^{\circ}$]{
\includegraphics[width=\figwidth]{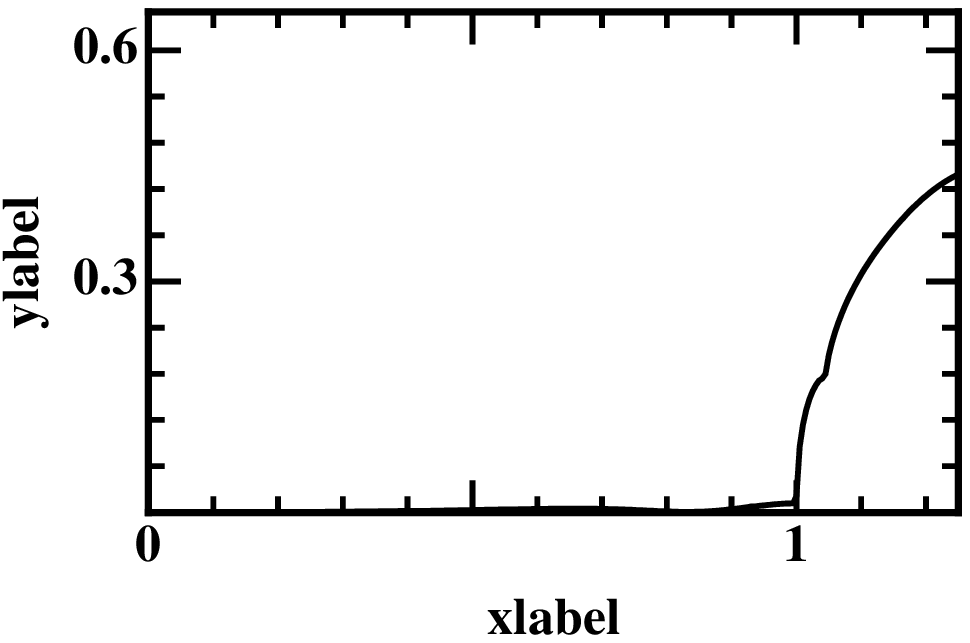}
}\\
\subfigure[180$^{\circ}$]{
\includegraphics[width=\figwidth]{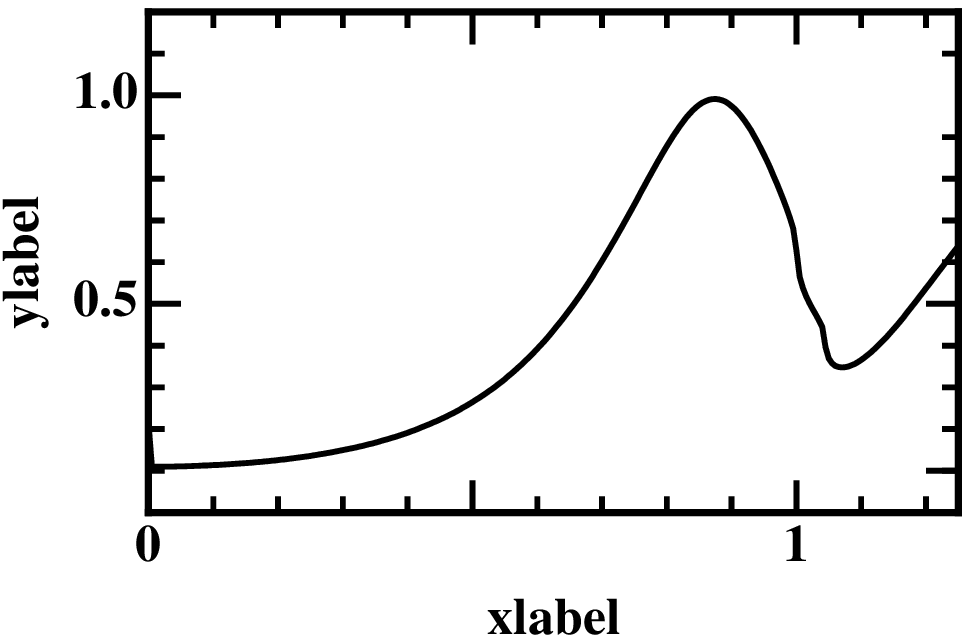}
}
\caption{
Conductance per spin for two leads at an angle of (a) 60$^{\circ}$,
(b) 120$^{\circ}$, and (c) 180$^{\circ}$ through a
hexagon with six $N$=20 armchair leads.
}\label{fig:Hex_AC_N=20_lowenergy}
\end{figure}
\end{psfrags}




%


\subsection{Zigzag Nanoribbon Junctions}

We conclude this section with a summary of analogous results for
a 120$^{\circ}$ zigzag nanoribbon junction.
Like armchair ribbons, zigzag ribbon widths may be characterized
by the minimum number of broken bonds $N$ required to sever
it, as illustrated in Fig. \ref{fig:ribbon_break}.
Unlike armchair ribbons,
zigzag ribbons are
metallic for any width \cite{brey1}.  At low energies the
current-carrying states have an interesting chirality in
that left- and right-moving states are associated with
different valley indices \cite{rycerz}.  It should be
emphasized that the association of this discrete index
can only be made precise in a continuum description.
All the geometries considered below may always be understood
in terms of semi-infinite ribbons joined together at
boundaries on the lattice scale.  Because the lowest
energy states of zigzag ribbons are highly confined to
the edges of the system, a pure continuum description
is inadequate even at the lowest energies.  Similar
physics has been noted recently in graphene
$p-n$ junctions \cite{akhmerov}.


This physics is most clearly seen in 120$^{\circ}$ zigzag junctions.
Fig. \ref{fig:120Bend_ZZ} illustrates the transmission for this geometry
for ribbon widths $N$=6,7, and 8.
At very low energies, there is a qualitative
difference between odd and even width ribbons, with the
former supporting a large conductance at zero energy and
the latter a small one.  Such odd-even behavior also occurs
in $p-n$ junctions, and appears to be related
to the fact that edges in a zigzag ribbon align when $N$
is even, but anti-align for odd $N$ \cite{akhmerov,waka}.
Even at higher energies, but still within the lowest
subband, the conductance as a function of energy appears
to change qualitatively from one width to the next
as $N$ increments by single units.

\begin{psfrags}
\psfrag{xlabel}[][][0.7]{$E_F/E_e$}
\psfrag{ylabel}[b][cl][0.7]{G $(e^2/h)$}
\begin{figure}[htb]
\centering
\subfigure[$N$=6]{
\includegraphics[width=\figwidth]{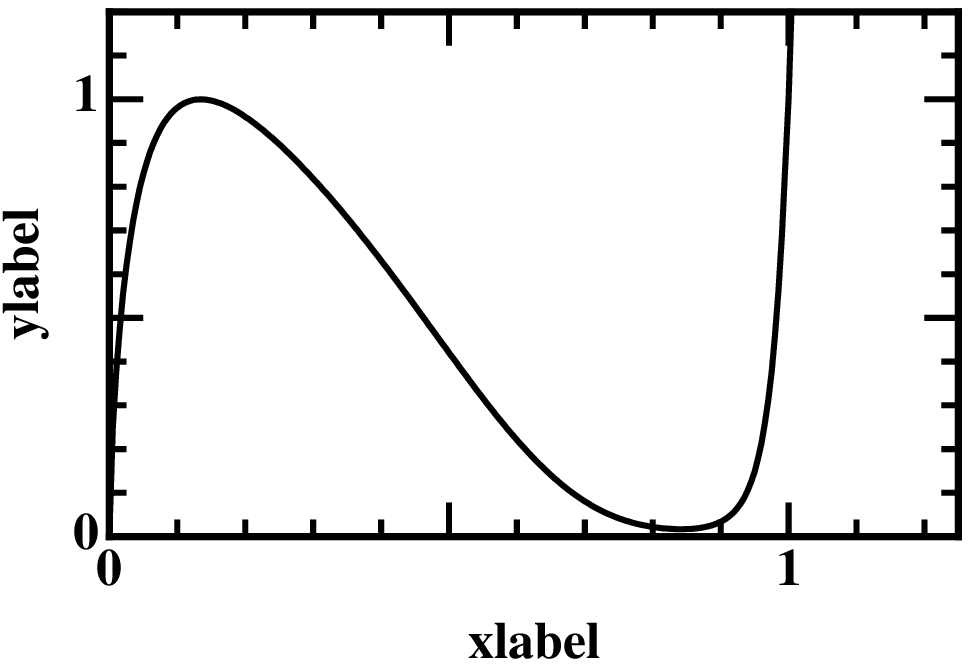}
}
\subfigure[$N$=7]{
\includegraphics[width=\figwidth]{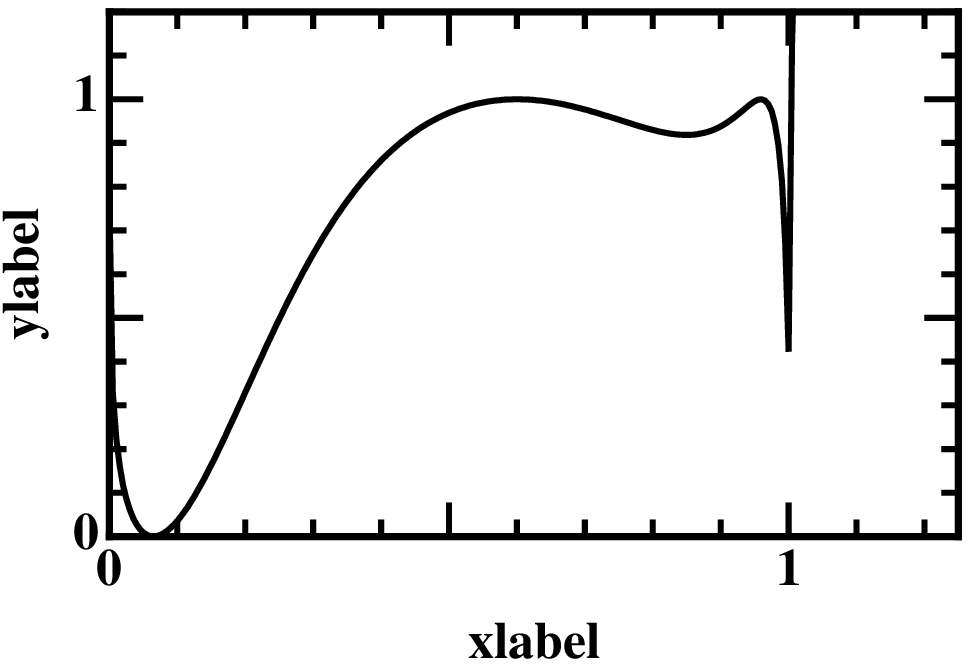}
}
\\
\subfigure[$N$=8]{
\includegraphics[width=\figwidth]{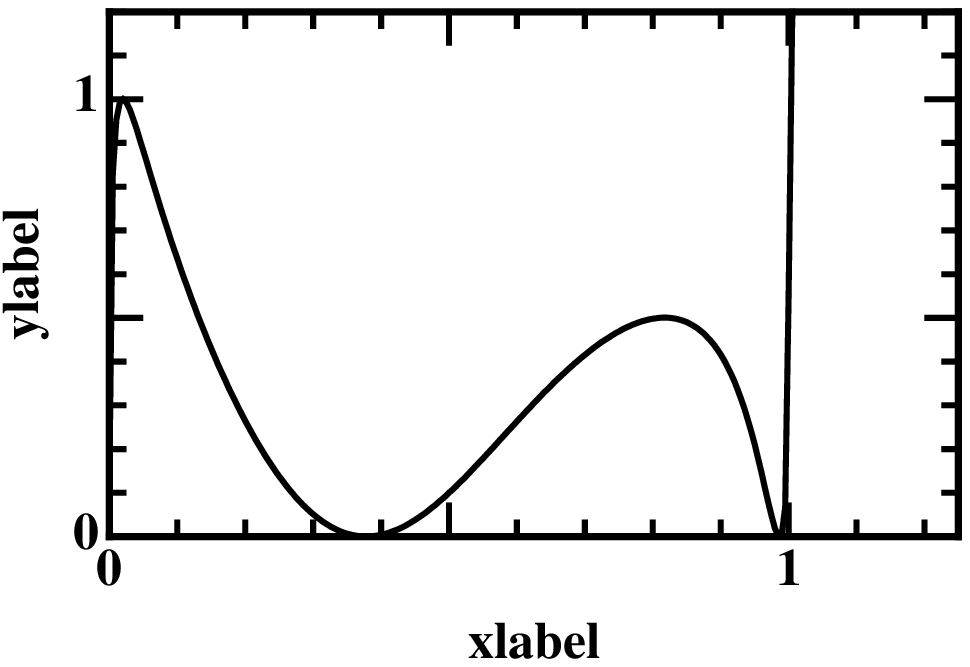}
}
\subfigure[Geometry for $N$=8]
{
\includegraphics[width=\figwidth]{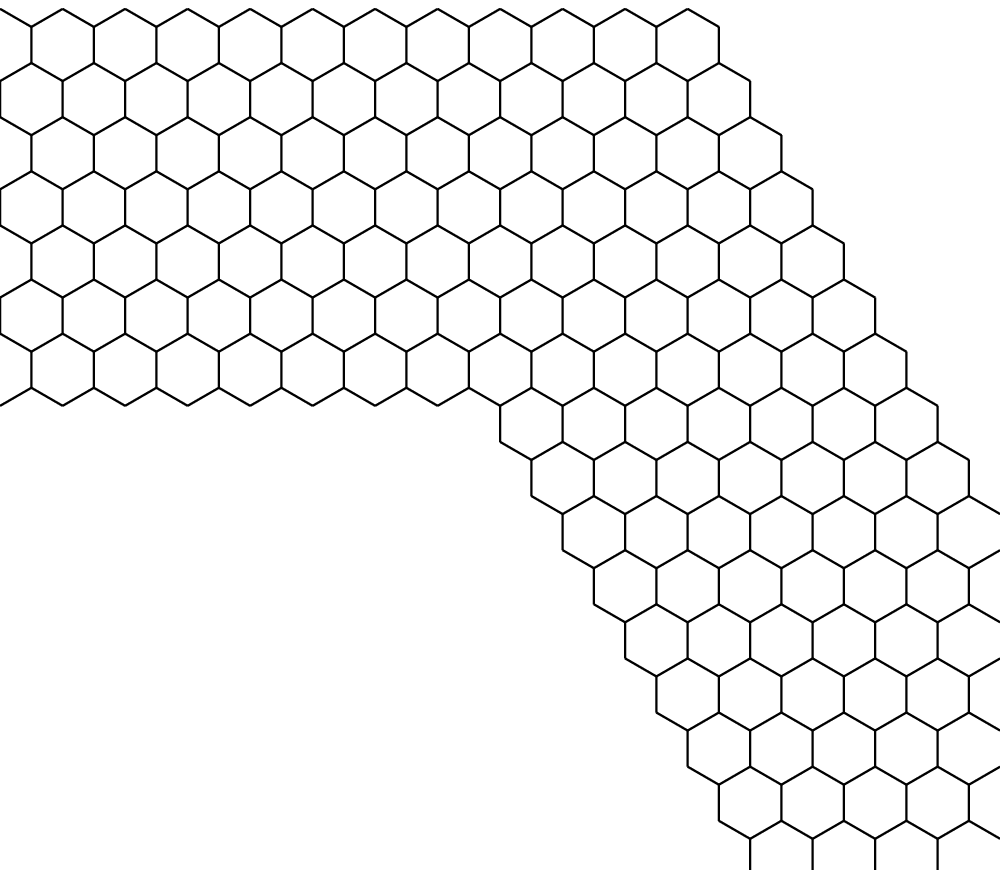}
}
\caption{Transmission through 120$^{\circ}$ bend for zigzag ribbons.
(a) $N$=6, (b) $N$=7, (c) $N$=8. (d) Geometry for $N$=8.
}
\label{fig:120Bend_ZZ}
\end{figure}
\end{psfrags}

\begin{figure}[t]
\vskip 1 cm
\centering
\subfigure[$E_F/E_e$=0.016]{
\includegraphics[width=\figwidth]{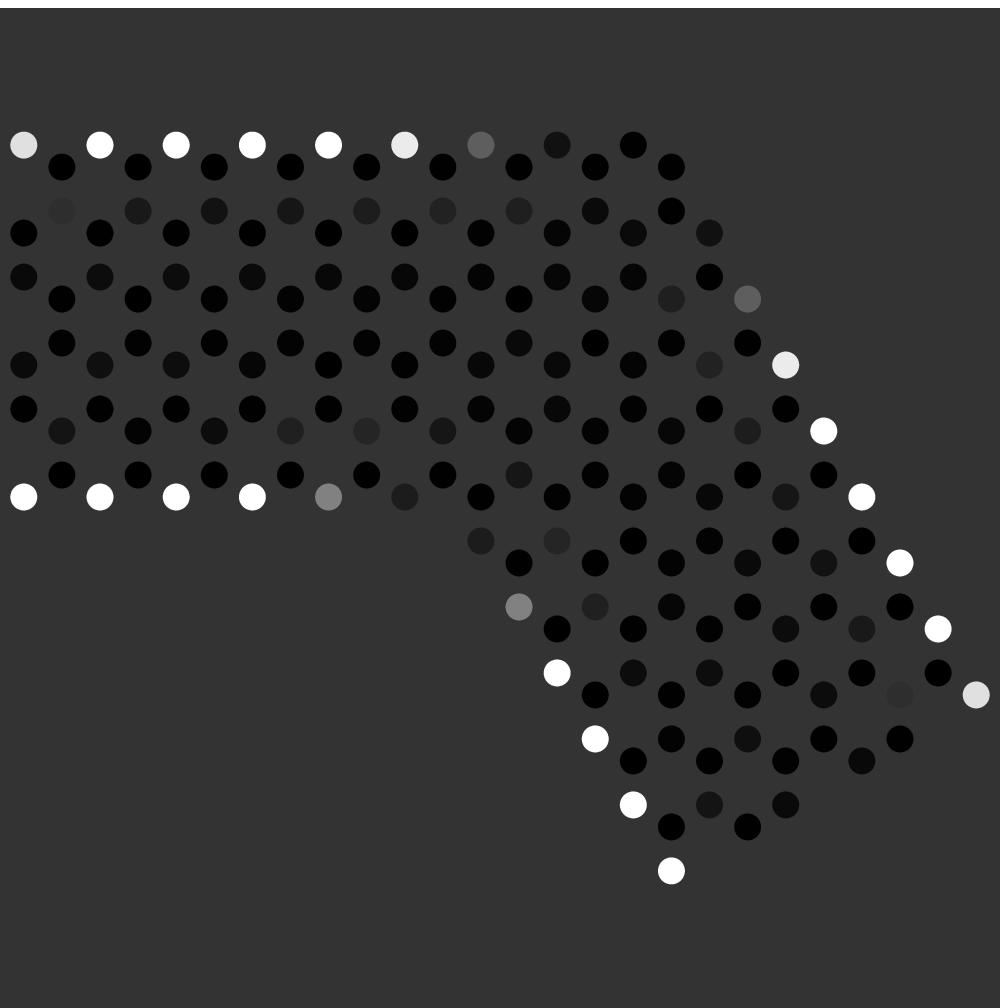}
}
\subfigure[$E_F/E_e$=0.16]{
\includegraphics[width=\figwidth]{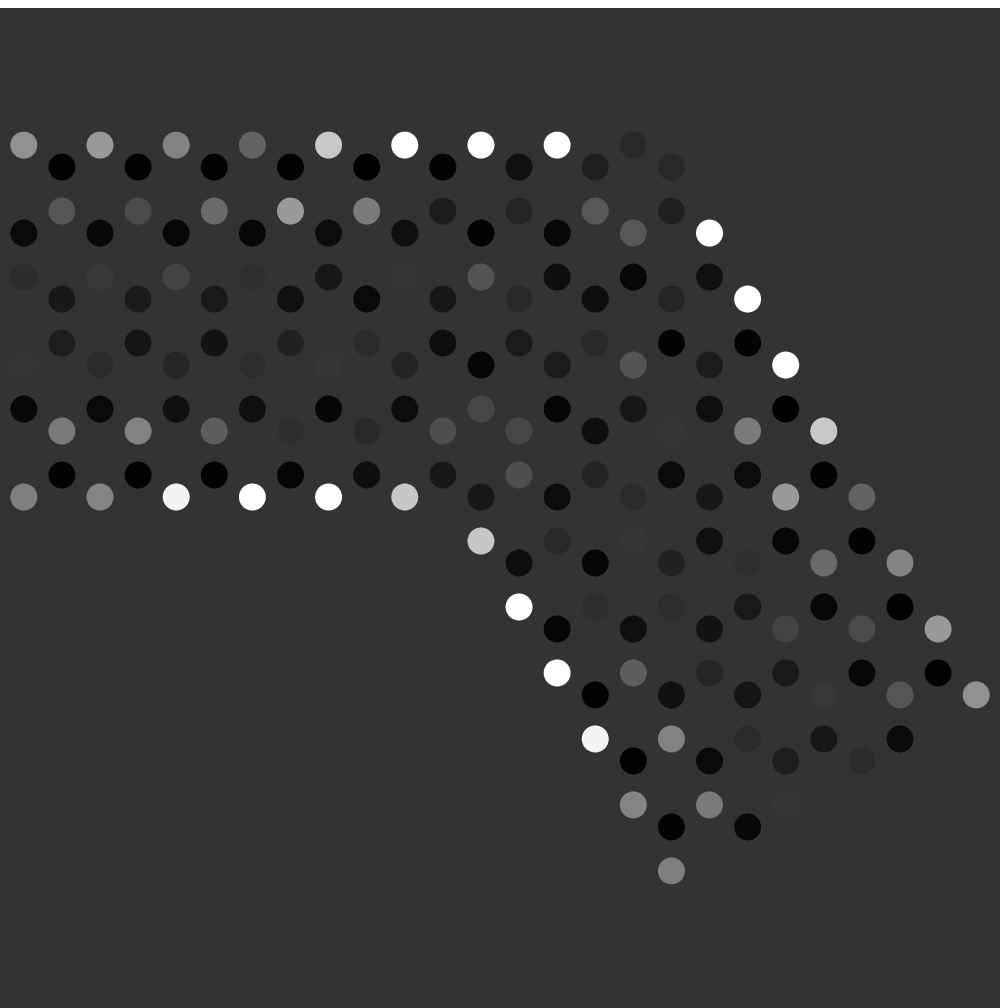}
}
\\
\subfigure[$E_F/E_e$=0.49]{
\includegraphics[width=\figwidth]{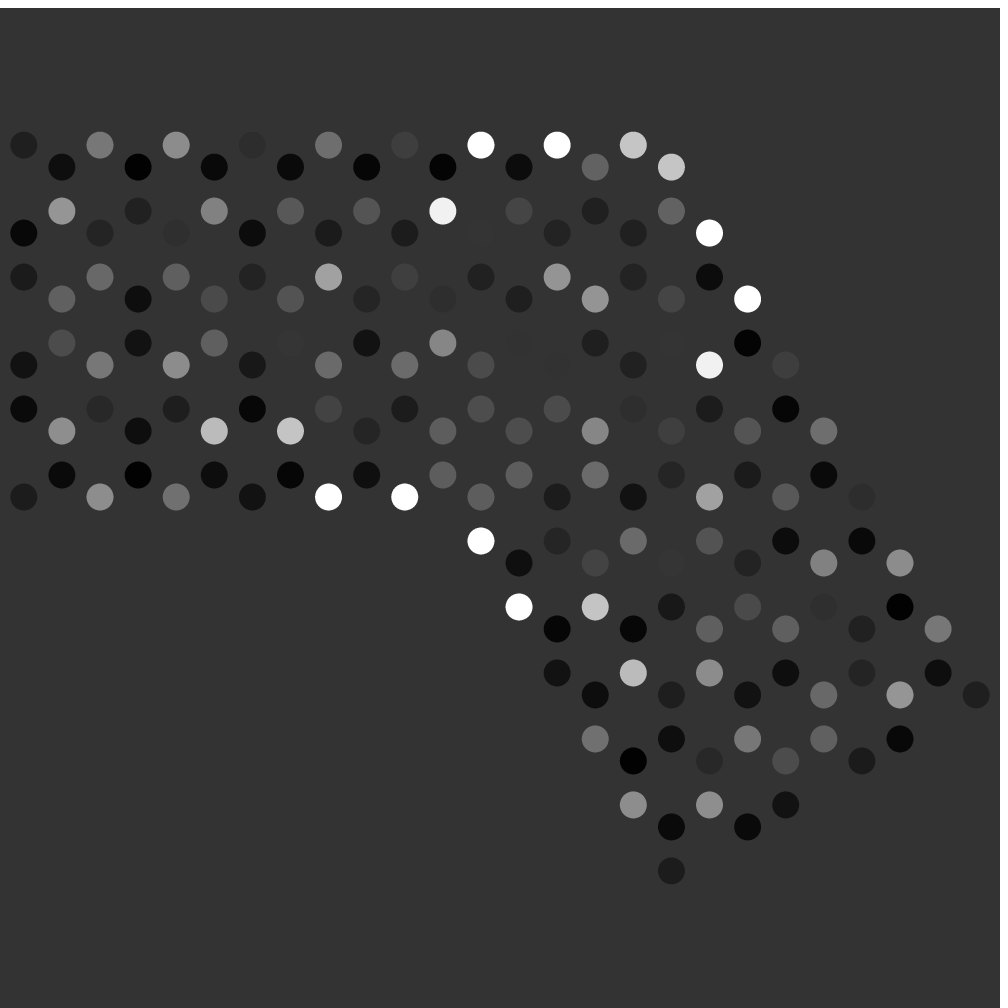}
}
\subfigure[$E_F/E_e$=0.82]{
\includegraphics[width=\figwidth]{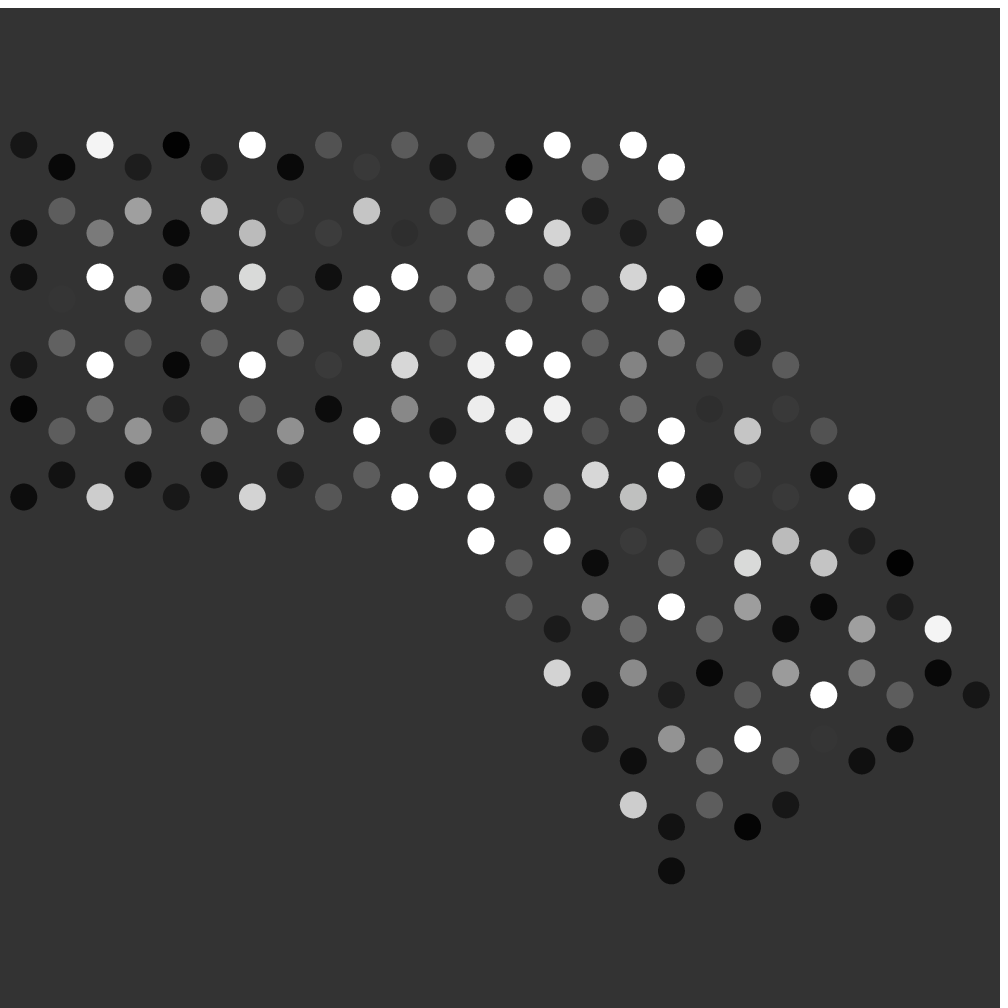}
}
\caption{
Local density of states shown in gray scale for 120$^{\circ}$ degree zigzag 
junction, $N=6$. 
Black denotes zero, and white denotes maximal LDOS. The leads modelled in the calculation 
extend away from the junction to infinity. }
\label{fig:120Bend_ZZ_LDOS}
\end{figure}

The importance of lattice scale physics in this system
is further made apparent by an examination of the local
density of states at low energies.  This is shown in
in Fig. \ref{fig:120Bend_ZZ_LDOS}
for a junction of width $N$=6 at several
energies below the first excited subband energy.
At low energies, while the wavefunction is strongly maximized
near the ribbon edges, the junction can attain zero or perfect
conductance, as is the case in (a) and (b), respsectively.
As the wavefunctions throughout the first
subband intimately involve the lattice scale, 
any continuum description for this system is not likely
to be reliable. 

\section{Conclusion}

In this paper
we have studied the conductance of various graphene
geometries in which the current in a nanoribbon  
is redirected through angles that
are multiples of 60$^{\circ}$, which can be accomplished
in the honeycomb network without introducing lattice defects or changing the 
ribbon type.
We focus on the low-energy behavior in armchair nanoribbon geometries 
and find
a variety of behaviors, including very high transmission for a
particular realization of a
simple 120$^{\circ}$ junction and
suppressed transmission for the same angle when there is an
intervening triangle.  A mode-matching analysis, within the
single-mode approximation (SMA), allows one to understand in a simple
way many of these results. With this technique we demonstrate that 
the rapid oscillation of the low
energy wavefunctions renders the conductance
through such junctions highly sensitive to the
precise way in which the ribbons are joined together.  Tight-binding
calculations support the conclusions of the SMA at low-energies,
and further elucidate the details of the conductance at
higher energies.

We also presented numerical results for conductance through other
geometries.  Hexagons in particular showed a dramatic suppression
of conductance through $120^{\circ}$, and further supported
a strong enhanced/suppressed conductance (depending on the angle
between the leads) at the van Hove singularity.
Zigzag nanoribbon junctions were also studied, and were
found to have a richer behavior than their armchair cousins,
which is likely to require a more microscopic description than
is possible in a continuum model.

\section{Acknowledgements} The authors thank J.P. Carini, F. Guinea and C. Tejedor for
useful discussions.  This work was supported by  MAT2006-03741 (Spain) (LB) 
and by the NSF through Grant No. DMR-0704033 (HAF).



\begin{thebibliography}{99}
\bibitem{com} Note that setting the wavefunctions equal on the A and B sublattices
for wavefunctions satisying the Dirac equation is essentially equivalent
to matching both wavefunctions and derivatives for those satisfying the
Schroedinger equation \cite{brey2}.
\bibitem{com2} Note that $p_0(\varepsilon)$ is actually imaginary
for very low energies, where the zigzag ribbon supports surface states.
\end{thebibliography}
\end{document}